\documentclass[11pt,preprint]{aastex}

\newcommand{\crra}{Cram\'er-Rao}

\slugcomment{To be submitted to PASP}

\shorttitle{Cram\'er \-Rao bound and Least-Squares in Astrometry}
\shortauthors{Lobos et al.}

\newtheorem{theorem}{\bf Theorem}
\newtheorem{preposition}{\bf Proposition}
\newtheorem{remark}{\bf Remark}
\begin{document}

\title{Performance analysis of the Least-Squares estimator in
  Astrometry}

\author{Rodrigo A. Lobos, Jorge F. Silva,}

\affil{Departamento de
  Ingenier\'{\i}a El\'ectrica, Facultad de Ciencias F\'{\i}sicas y
  Matem\'aticas,\\Universidad de Chile, Beauchef 850, Santiago, Chile}
\email{rlobos, josilva @ing.uchile.cl}

\author{Rene A. Mendez,}
\affil{Departamento de Astronom\'{\i}a, Facultad de Ciencias
  F\'{\i}sicas y Matem\'aticas,\\Universidad de Chile, Casilla 36-D,
  Santiago, Chile} \email{rmendez@u.uchile.cl}

\and

\author{Marcos Orchard}
\affil{Departamento de
  Ingenier\'{\i}a El\'ectrica, Facultad de Ciencias F\'{\i}sicas y
  Matem\'aticas,\\Universidad de Chile, Beauchef 850, Santiago, Chile}
\email{morchard@ing.uchile.cl}

\clearpage

\begin{abstract}
We characterize the performance of the widely-used least-squares
estimator in astrometry in terms of a comparison with the \crra\ lower
variance bound. In this inference context the performance of the
least-squares estimator does not offer a closed-form expression, but a
new result is presented (Theorem~\ref{ls_performances_bounds}) where
both the bias and the mean-square-error of the least-squares estimator
are bounded and approximated analytically, in the latter case in terms
of a {\em nominal value} and an interval around it. From the predicted
nominal value we analyze how efficient is the least-squares estimator
in comparison with the minimum variance \crra\ bound. Based on our
results, we show that, for the high signal-to-noise ratio regime, the
performance of the least-squares estimator is significantly poorer
than the \crra\ bound, and we characterize this gap analytically. On
the positive side, we show that for the challenging low
signal-to-noise regime (attributed to either a weak astronomical
signal or a noise-dominated condition) the least-squares estimator is
near optimal, as its performance asymptotically approaches the
\crra\ bound. However, we also demonstrate that, in general, there is
no unbiased estimator for the astrometric position that can precisely
reach the \crra\ bound. We validate our theoretical analysis through
simulated digital-detector observations under typical observing
conditions. We show that the {\em nominal value} for the
mean-square-error of the least-squares estimator (obtained from our
theorem) can be used as a benchmark indicator of the expected
statistical performance of the least-squares method under a wide range
of conditions. Our results are valid for an idealized linear
(one-dimensional) array detector where intra-pixel response changes
are neglected, and where flat-fielding is achieved with very high
accuracy.
\end{abstract}


\clearpage

\section{Motivation} \label{sec_intro}


Astrometry, that branch of observational astronomy that deals with the
precise and accurate estimation of angular positions of light-emitting
(usually point-like) sources projected against the celestial sphere,
is the oldest technique employed in the study of the heavens
\citep{hog2009, reffert2009, hog2011}. Repeated measurements of
positions, spread over time, allow a determination of the motions and
distances of theses sources, with astrophysical implications on
dynamical studies of stellar systems and the Milky Way as a
whole. With the advent of solid-state detectors and all-digital
techniques applied to radio-interferometry and specialized ground- and
space-based missions, astrometry has been revolutionized in recent
years, as we have entered a high-precision era in which this technique
has started to play an increasingly important role in all areas of
astronomy, astrophysics \citep{vanaltena2013}, and cosmology
\citep{lattanzi2012}.

Current technology, based on two-dimensional discrete digital
detectors (such as {\em charged coupled devices} - CCDs), record a
(noisy) image (on an array of photo-sensitive pixels) of celestial
sources, from which it is possible to estimate both their astrometry
and photometry, simultaneously \citep{howell2006handbook}. The
inference problem associated to the determination of these quantities
is at the core of the astrometric endeavor described previously.


A number of techniques have been proposed to estimate the location and
flux of celestial sources as recorded on digital detectors. In this
context, estimators based on the use of a least-squares error function
(LS hereafter) have been widely adopted
\citep{stetson1987daophot,king1983accuracy,alard1998method,honeycutt1992ccd,cameron2006fast}. The
use of this type of decision rule has been traditionally justified
through heuristic reasons and because they are conceptually
straightforward to formulate based on the observation model of these
problems. Indeed, the LS approach was the classical method used when
the observations were obtained with analog devices
\citep{vanAltena_1975,euer1978} (which corresponds to a Gaussian noise
model for the observations, different from that of modern digital
detectors, which is characterized instead by a Poisson statistics)
and, consequently, the LS method was naturally adopted from the
analogous to the digital observational setting\footnote{See footnote
  11 in Section~\ref{subsec_achie}.}.

In contemporary astrometry (Gaia, for instance), stellar positions
will be obtained by optimizing a likelihood function (see, e.g.,
\citet{lind2000}, which uses the equivalent of our
equations~(\ref{eq_pre_5}) and~(\ref{log-like}) in
Sections~\ref{sub_sec_astro_photo} and~\ref{subsec_achie}
respectively), not by LS. Nevertheless, since LS methods offer
computationally efficient implementations and have shown reasonable
performance \citep{lee1983theoretical,stone1989} they are still widely
used in astrometry either on general-purpose software packages for the
analysis of digital images such as DAOPHOT \citep{stetson1987daophot},
or on dedicated pipelines, such as that adopted in the Sloan Digital
Sky Survey survey (SDSS hereafter,
\citet{luptonetal2001,lupton2007}). For example, in DAOPHOT astrometry
(and photometry) are obtained through a two-step process which
involves a LS minimization of a trial function (e.g., a bi-dimensional
Gaussian, see \citet[equation~6]{stetson1987daophot}, equivalent to
our one-dimensional case in equation~(\ref{eq_subsec_mse_of_LS_1}),
Section~\ref{subsec_achie}), and then applying a correction by means
of an empirically determined look-up table (also computed performing a
LS on a set of high signal-to-noise ratio images distributed over the
field of view of the image, see
\citet[equation~8]{stetson1987daophot}). This last step accounts for
the fact that the PSF of the image under analysis may not be exactly
Gaussian. The SDSS pipeline (\citet{luptonetal2001}) obtains its
centroids also through a two-step process: First it fits a
Karhunen-Lo\`eve transform (KL transform hereafter\footnote{Also
  referred to as ``principal component analysis'', ``proper orthogonal
  decomposition'', ``empirical orthogonal functions'' (a term used in
  meteorology and geophysics), or the Hotelling transform, see
  \citet{graydavi10}.}, see, e.g., \citet{najim2008}) to a set of
isolated bright stars in the field-of-view of the image, and then it
uses the base functions determined in this way, to fit the astrometry
and photometry for the object(s) under consideration using a LS
minimization scheme (see \citet[equation~(5)]{lupton2007}). Both
codes, DAOPHOT and the SDSS pipeline have been extensively used and
tested by the astronomical community, giving very reliable results
(see, e.g., \citet{pieret03,becket07}).


Considering that LS methods are still in use in astrometry, and driven
by the increase in the intrinsic precision available by the new
detectors and instrumental settings, and by the fact that CCDs will
likely continue to be the detector of choice for the focal-plane in
science-quality imaging application at optical wavelengths for both
space- as well as ground-borne programs\footnote{See, e.g.,
  http://www.bnl.gov/cosmo2013/}, it is timely to re-visit the
pertinence of the use of LS estimators. Indeed, in the digital
setting, where we observe discrete samples (or counts) on a photon
integrating device, there is no formal justification that the LS
approach is optimal in the sense of minimizing the mean-square-error
(MSE) of the parameter estimation, in particular for astrometry, which
is the focus of this work.


The question of optimality (in some statistical sense) has always been
in the interest of the astronomical community, in particular the idea
of characterizing fundamental performance bounds that can be used to
analyze the efficiency of the adopted estimation schemes. In this
context, we can mention some seminal works on the use of the
celebrated \crra\ (CR hereafter) bound in astronomy by
\citet{lindegren1978,jakobsen1992,zaccheo1995,adorf1996}; and
\citet{bastian2004}.  The CR bound is a minimum variance (MV) bound
for the family of unbiased estimators
\citep{radhakrishna1945information,cramer1946contribution}. In
astrometry and photometry this bound has offered meaningful
closed-form expressions that can be used to analyze the complexity of
the inference task, and its dependency on key observational and design
parameters such as the position of the object in the array, the
intensity of the object, the signal-to-noise ratio (SNR hereafter),
and the resolution of the instrument
\citep{Winick:86,1989perryman,lindegren2010,2013mendez,2014mendez}.
In particular, for photometry, \citet{1989perryman} used the CR bound
to show that the LS estimator is a good estimator, achieving a
performance close to the limit in a wide range of observational
regimes, and approaching very closely the bound at low SNR. In
astrometry, on the other hand, \citet{2013mendez,2014mendez} have
recently studied the structure of this bound and have analyzed its
dependency with respect to important observational parameters, under
realistic astronomical observing conditions. In those works,
closed-form expressions for the bound were derived in a number of
important settings (high spatial resolution, low and high SNR), and
their trends were explored across angular resolution and the position
of the object in the array.  As an interesting outcome, the analysis
of the CR bound allows us to find the optimal pixel resolution of the
array for a given setting, as well as providing formal justification
to some heuristic techniques commonly used to improve performance in
astrometry, like {\em dithering} for undersampled images
\citep[Sec. 3.3]{2013mendez}.

The specific problem of evaluating the existence of an estimator that
achieves the CR bound has not been covered in the literature, and
remains an interesting open problem. On this, \citet{2013mendez} have
empirically assessed (using numerical simulations) the performance of
two LS methods and the maximum-likelihood (ML hereafter) estimator,
showing that their variances follow very closely the CR limit in some
specific regimes. In this paper, we analyze in detail the performance
of the LS estimator with respect to the CR bound, with the goal of
finding concrete regimes, if any, where this estimator approaches the
CR bound and, consequently, where it can be considered an efficient
solution to the astrometric problem. This application is a challenging
one, because estimators based on a LS type of objective function do
not have a closed-form expression in astrometry. In fact, this
estimation approach corresponds to a non linear regression problem,
where the resulting estimator is implicitly defined. As a result, no
expressions for the performance of the LS estimator can be obtained
analytically.
To address this issue, our main result (Theorem
\ref{ls_performances_bounds}, Section~\ref{subsec_mse_of_LS}) derives
expressions that bound and approximate the variance of the LS
estimator. Our approach is based on the work by \citet{dsp2013}, where
the authors tackle the problem of approximating the bias and MSE of
general estimators that are the solution of an optimization
problem. In \citet{fessler1996} another methodology is given to
approximate the variance and mean of implicitly defined estimators,
which has been applied to medical imaging and acoustic source
localization \citep{raykar2005}.

The main result of our paper is a refined version of the result
presented in \citet{dsp2013}, where one of their key assumptions,
which is not applicable in our estimation problem, is reformulated. In
this process, we derive lower and upper bounds for the MSE performance
of the LS estimator. Using these bounds, we analyze how closely the
performance of the LS estimator approaches the CR bound across
different observational regimes. We show that for high SNR there is a
considerable gap between the CR bound and the performance of the LS
estimator. Remarkably, we show that for the more challenging low SNR
observational regime (weak astronomical sources), the LS estimator is
near optimal, as its performance is arbitrarily close to the CR bound.

The paper is organized as follows. Section \ref{sec_pre} introduces
the problem, notation, as well as some preliminary results.  Section
\ref{main_sec} represents the main contribution, where Theorem
\ref{ls_performances_bounds} and its interpretation are introduced.
Section \ref{subsec_empirical} shows numerical analyses of the
performance of LS estimator under different observational regimes.
Finally Section \ref{final} provides a summary of our results, and
some final remarks.


\section{Introduction and preliminaries} \label{sec_pre}

In this section we introduce the problem of astrometry
as well as concepts and definitions that will be used throughout the
paper. For simplicity, we focus on the 1-D scenario of a linear array
detector, as it captures the key conceptual elements of the
problem\footnote{This analysis can be extended to the 2-D case as
  shown in \citet{2013mendez}.}.

\subsection{Astrometry and photometry based on a photon integrating device} \label{sub_sec_astro_photo}

The specific problem of interest is the inference of the position of a
point source. This source is parameterized by two scalar quantities,
the position of the object $x_c\in \mathbb{R}$ in the
array\footnote{This is of course related to an angular position in the
  sky, measured in arcsec, through the ``plate-scale'', which is an
  optical design feature determined by the instrument plus telescope
  configuration.}, and its intensity (or brightness, or flux) that we
denote by $\tilde{F}\in \mathbb{R}^+$.  These two parameters induce a
probability $\mu_{x_c,\tilde{F}}$ over an observation space that we
denote by $\mathbb{X}$.  More precisely, given a point source
represented by the pair $(x_c,\tilde{F})$, it creates a nominal
intensity profile in a photon integrating device, typically a CCD,
which can be generally written as:

\begin{equation}\label{eq_pre_1}
	\tilde{F}_{x_c, \tilde{F}}(x)=\tilde{F} \cdot
\phi(x-x_c,\sigma),
\end{equation}

where $\phi(x-x_c,\sigma)$ denotes the one dimensional normalized
point spread function (PSF) evaluated on the pixel coordinate $x-x_c$,
and where $\sigma$ is a generic parameter that determines the width
(or spread) of the light distribution on the detector (typically a
function of wavelength and the quality of the observing site, see
Section~\ref{subsec_empirical}) (see \citet{2013mendez,2014mendez} for
more details).

The profile in equation~(\ref{eq_pre_1}) is not observed directly, but
through three sources of perturbations: First, an additive background
which accounts for the photon emissions of the open (diffuse) sky and
the contributions from the noise of the instrument itself (the
read-out noise and dark-current \citep{gilligand1992,tyson1986})
modeled by $\tilde{B}_i$ in equation~(\ref{eq_pre_2b}). Second, an
intrinsic uncertainty between the aggregated intensity (the nominal
object brightness plus the background) and the actual measurements,
denoted by $I_i$ in what follows, which is modeled by independent
random variables that obey a Poisson probability law. And, finally, we
need to consider the spatial quantization process associated with the
pixel-resolution of the detector as specified by $g_i(x_c)$ in
equations~(\ref{eq_pre_2b}) and~(\ref{eq_pre_3})\footnote{Note
    that pixel-convolved point-spread function $g_i(x_c)$ is sometimes
    referred to as the the "pixel response function".}. Including
these three effects, we have a countable collection of independent and
not identically distributed random variables (observations or counts)
$\left\{I_i: i \in \mathbb{Z}\right\}$, where $I_i \sim
Poisson(\lambda_i(x_c, \tilde{F}))$, driven by the expected intensity
at each pixel element $i$, given by:
\begin{equation}\label{eq_pre_2b}
	\lambda_i(x_c, \tilde{F}) \equiv\mathbb{E}\{I_i\}= \underbrace{\tilde{F} \cdot
          g_i(x_c)}_{\equiv \tilde{F}_i(x_c,\tilde{F})} + \tilde{B}_i,~\forall i\in \mathbb{Z}
\end{equation}
and,
\begin{equation}\label{eq_pre_3}
	g_i(x_c) \equiv \int^{x_i+\Delta x/2}_{x_i-\Delta x/2} \phi(x- x_c,\sigma) \, dx, \ \forall i \in \mathbb{Z},
\end{equation}
where $\mathbb{E}$ represents the expectation value of the argument,
and $\left\{x_i: i \in \mathbb{Z}\right\}$ denotes the standard
uniform quantization of the real line-array with resolution $\Delta
x>0$, i.e., $x_{i+1}-x_i=\Delta x$ for all $i \in \mathbb{Z}$. In
practice, the detector has a finite collection of measured elements
(or pixels) $x_1,...,x_n$, then a basic assumption here is that we
have a good coverage of the object of interest, in the sense that for
a given position $x_c$:
%
\begin{equation}\label{eq_pre_4}
	\sum_{i=1}^n g_i(x_c) \approx \sum_{i\in \mathbb{Z}} g_i(x_c)
        =\int_{-\infty}^{\infty} \phi(x-x_c,\sigma) \, dx = 1.
\end{equation}
Note that equation~(\ref{eq_pre_3}) adopts the idealized
  situation where every pixel has the exact same response function
  (equal to unity), or, equivalently, that our flat-field process has
  been achieved with minimal uncertainty. It also assumes that the
  intra-pixel response is uniform. The latter is more important in the
  severely undersampled regime (see, e.g.,
  \citet[Figure~1]{adorf1996}) which is not explored in this
  paper. However a relevant aspect of data calibration is achieving a
  proper flat-fielding which can affect the correctness of our
  analysis and the form of the adopted likelihood function (see
  below).

At the end, the likelihood of the joint random observation vector
$I^n=(I_1,..,I_n)$ (with values in $\mathbb{N}^n$), given the source
parameters $(x_c, \tilde{F})$, is given by:
\begin{equation}\label{eq_pre_5}
L(I^n;x_c,\tilde{F}) = f_{\lambda_1(x_c,\tilde{F})}(I_1) \cdot
f_{\lambda_2(x_c,\tilde{F})}(I_2) \cdots
f_{\lambda_n(x_c,\tilde{F})}(I_n), \ \forall I^n\in \mathbb{N}^n,
\end{equation}
where $f_{\lambda}(x)=\frac{e^{-\lambda}\cdot \lambda^x}{x!}$ denotes
the probability mass function (pmf) of the Poisson law
\citep{graydavi10}. We emphasize that equation~(\ref{eq_pre_5})
  assumes that the observations (contained in the individual pixels,
  denoted by the index $i$) are independent (although not identically
  distributed, since they follow $\lambda_i$). Of course, this is only
  an approximation to the real situation since it implies, in
  particular, that we are neglecting any electronic defects or
  features in the device such as, e.g., the cross-talk present in
  multi-port CCDs \citep{freyetal01}, or read-out correlations such as
  the odd-even column effect in IR detectors \citep{mason07}, as well
  as calibration or data reduction deficiencies (e.g., due to
  inadequate flat-fielding \citep{gawiet06}) that may alter this
  idealized detection process. A serious attempt is done by
  manufacturers and observatories to minimize the impact of these
  defects, either by an appropriate electronic design, or by adjusting
  the detector operational regimes (e.g., cross-talk can be reduced to
  less than 1 part in $10^4$ by adjusting the readout speed and by a
  proper reduction process, \citet{freyetal01}). In essence, we are
  considering an ideal detector that would satisfy the proposed
  likelihood function given by equation~(\ref{eq_pre_5}), in real
  detectors the likelihood function could be considerably more
  complex\footnote{Actually, in a real situation we may not even be
    able to write such a function due to our imperfect
    characterization or limited knowledge of the detector device.}.

We can formulate the astrometric and photometric estimation task, as
the problem of characterizing a decision rule $\tau_n():\mathbb{N}^n
\rightarrow \Theta$, with $\Theta=\mathbb{R}^2$ being a parameter
space, where given an observation $I^n$ the parameters to be estimated
are $(\hat{x}_c(I^n),\hat{\tilde{F}}(I^n))=\tau_n(I^n)$. In other
words, $\tau_n(I^n)$ gives us a prescription (or statistics) that
would allow us to estimate the underlying parameters $(x_c,\tilde{F})$
(the estimated parameters are denoted by
$(\hat{x}_c,\hat{\tilde{F}})$) from the available data vector $I^n$.

In the simplest scenario, in which one is interested in determining a
single (unknown) parameter $\theta$ (e.g., in our case either $x_c$ or
$\tilde{F}$, assuming that all other parameters are perfectly well
known), a commonly used decision rule adopted in statistics to
estimate this parameter (the estimation being $\hat{\theta}$) is to
consider the prescription of minimum variance (denoted by $Var$),
given by:

\begin{eqnarray}\label{minvar}
\hat{\theta}() & \equiv & \arg \min_{\alpha} Var
(\alpha(I^n)) \nonumber \\
& = & \arg \min_{\alpha}  \mathbb{E}_{I^n \sim
  f_{\theta}^n} \left( \alpha (I^n) - \theta \right)^2
\end{eqnarray}

where ``$ \arg \min$'' represents the argument that minimizes the
expression, while $\alpha$ is a generic variable representing the
parameter to be determined. Note that in the last equality we have
assumed that $\hat \theta()$ is an unbiased estimator of the parameter
(i.e., that $\mathbb{E}(\hat\theta)= \theta$), so that under this rule
we are implicitly minimizing the MSE of the estimate with respect to
the hidden true parameter $\theta$.

Unfortunately, the general solution of equation~(\ref{minvar}) is
intractable, as in principle it requires the knowledge of $\theta$,
which is the essence of the inference problem. An additional issue
with equation~(\ref{minvar}) is that, by itself, it does not provide
an analytical expression that tells us how to compute $\hat \theta$ in
terms of $I^n$\footnote{In many cases one can propose an ``educated
  guess'' for $\hat\theta$ (e.g., a function of some sort) which is
  trained with a (``good'') subset of the data itself (thus
  approximately removing the ambiguity that the true parameter
  $\theta$ is, in fact, unknown). This heuristic approach is adopted,
  e.g., in the SDSS pipeline trough the use of the KL transform, which
  is a good approximation to a matched filter (also known as the
  ``North filter'', see e.g., \citet{graydavi10}).}. Fortunately,
there are performance bounds that characterize how far can we be from
the theoretical solution in equation~(\ref{minvar}), and even
scenarios where the optimal solution can be achieved in a closed-form
(see Section~\ref{subsec_achie} and
Appendix~\ref{proof_pro_achie_astrometry}).  One of the most
significant results in this field is the CR minimum variance bound
which will be further explained below.


\subsection{The \crra\ bound} \label{subsec_cr_bounds}
The CR bound offers a performance bound on the variance of the family
of unbiased estimators.
More precisely, let $\left\{ I_i \right\} _{i=1}^n$ be a collection of
independent observations that follow a parametric pmf
$f_{\bar{\theta}}$ defined on $\mathbb{N}$. The parameters to be
estimated from $( I_1,I_2,...,I_n)$
will be denoted in general by the vector
$\bar{\theta}=(\theta_1,\theta_2,...,\theta_m) \in \Theta =
\mathbb{R}^m$.  Let $\tau_n(I^n):\mathbb{N}^n \rightarrow \Theta$ be
an unbiased estimator\footnote{In the sense that, for all
  $\bar{\theta}\in \Theta$, $\mathbb{E}_{I^n \sim f^n_{\bar{\theta}}}
  \left\lbrace \tau_n(I^n)\right\rbrace =\bar{\theta}$.}  of
$\bar{\theta}$, and $L(I^n;\bar{\theta})= f_{\bar{\theta}}(I_1)\cdot
f_{\bar{\theta}}(I_2) \cdots f_{\bar{\theta}}(I_n) $ be the likelihood
of the observation
$I^n \in \mathbb{N}^n$ given $\bar{\theta}\in \Theta$.  Then, the CR
bound \citep{radhakrishna1945information,cramer1946contribution}
establishes that if:
%
\begin{equation} \label{cond2d}
\mathbb{E}_{I^n \sim f^n_{\bar{\theta}}}\left\lbrace \frac{\partial \ln
  L(I^n; \bar{\theta}) }{\partial \theta_i} \right\rbrace = 0, \;\;  \forall i \in \left\{1,\ldots, m \right\},
\end{equation}
then,
\begin{equation}\label{varcr}
Var (\tau_n(I^n)_i) \geq [
  \mathcal{I}_{\bar{\theta}}(n)^{-1} ]_{i,i},
\end{equation}
where $\mathcal{I}_{\bar{\theta}}(n)$ is the {\em Fisher information}
matrix given by:
\begin{equation}\label{fisher}
[ \mathcal{I}_{\bar{\theta}}(n)]_{i,j} = \mathbb{E}_{I^n \sim f^n_{\bar{\theta}}} \left\lbrace
\frac{\partial \ln L(I^n; \bar{\theta})}{\partial \theta_i} \cdot
\frac{\partial \ln L(I^n; \bar{\theta}) }{\partial \theta_j}
\right\rbrace \;\; \forall i,j \in  \left\{1,\ldots, m \right\}.
\end{equation}
In particular, for the scalar case ($m=1$), we have that for all
$\theta \in \Theta$:
\begin{equation}\label{cr_scalar}
	\min_{\tau_n(\cdot)\in \mathcal{T}^n} Var(\tau_n(I^n)) \geq \mathcal{I}_\theta(n)^{-1}=  \mathbb{E}_{I^n \sim f^n_{\theta}}  \left\lbrace\left[  \left(\frac{d \ln L(I^n; {\theta})}{d \theta} \right)^2 \right]\right\rbrace^{-1},
\end{equation}
where $\mathcal{T}^n$ is the collection of unbiased estimators and
$I^n \sim f^n_\theta$.

%
Returning to our problem in Section \ref{sub_sec_astro_photo},
\citet{2013mendez,2014mendez} have characterized and analyzed the CR
bound for the isolated problem of astrometry and photometry,
respectively, as well as the joint problem of photometry and
astrometry.  Particularly, we highlight the following results, which
will be used later on:

\begin{preposition} (\citet[pp. 800]{2014mendez}) 
	\label{pro_FI_photometry_astrometry}
	Let us assume that $x_c \in \mathbb{R}$ is fixed and known,
        and we want to estimate $\tilde{F}$ (fixed but unknown) from
        $I^n \sim f_{(x_c,\tilde{F})} $ in equation~(\ref{eq_pre_5}).  In this
        scalar parametric context, the Fisher information is given by:
	\begin{equation}\label{fi_photometry}
		\mathcal{I}_{\tilde{F}}(n) = \sum_{i=1}^n
                \frac{g_i(x_c)}{\tilde{F} g_i(x_c) + \tilde{B}_i},
	\end{equation}
	which from equation~(\ref{cr_scalar}) induces a MV bound for
        the {\em photometry estimation problem}.  On the other hand,
        if $\tilde{F} \in \mathbb{R}^+$ is fixed and known, and we
        want to estimate $x_c$ (fixed but unknown) from $I^n \sim
        f_{(x_c,\tilde{F})} $ in equation~(\ref{eq_pre_5}), then the
        Fisher information is given by:
	\begin{equation}\label{fi_astrometry}
		\mathcal{I}_{x_c}(n) = \sum_{i=1}^n \frac{ \left(
                  \tilde{F}\frac{d g_i(x_c)}{d x_c} \right)^2
                }{\tilde{F} g_i(x_c) + \tilde{B}_i} \equiv \frac{1}{\sigma_{CR}^2},
	\end{equation}
	which from equation~(\ref{cr_scalar}) induces a MV bound for
        the {\em astrometric estimation problem}, and where
        $\sigma_{CR}^2 \equiv \mathcal{I}_{x_c}(n)^{-1}$ denotes the
        (astrometric) CR bound.
\end{preposition}

At this point it is relevant to study if there is any practical
estimator that achieves the CR bound presented in
equations~(\ref{fi_photometry}) and~(\ref{fi_astrometry}) for the
photometry and astrometry problem, respectively.  For the photometric
case, \citet[their Appendix A]{1989perryman} has shown that the
classical LS estimator is near-optimal, in the sense that its variance
is close to the CR bound for a wide range of experimental regimes, and
furthermore, in the low SNR regime, when $\tilde{F}g_i(x_c) \ll
\tilde{B}_i$, its variance (determined in closed-form) asymptotically
achieves the MV bound $\mathcal{I}_{\tilde{F}}(n)^{-1}$ in
equation~(\ref{fi_photometry}).  This is a formal justification for
the goodness of the LS as a method for doing isolated photometry in
the setting presented in Section \ref{sub_sec_astro_photo}. An
equivalent analysis has not been conducted for the astrometric
problem, which is the focus of the next section of this work.

\section{Achievability analysis of the \crra\ bound in astrometry}
\label{main_sec}

We first evaluate if the CR bound for the astrometric problem,
$\sigma_{CR}^2$ from equation~(\ref{fi_astrometry}), can possibly be
achieved by any unbiased estimator. Then, we focus on the widely used
LS estimation approach, to evaluate its performance in comparison with
the astrometric MV bound presented in Proposition
\ref{pro_FI_photometry_astrometry}.

\subsection{Non achievability} \label{subsec_achie}
Concerning achievability, we demonstrate that for astrometry (i.e.,
assuming $\tilde{F}$ is known) there is no estimator that achieves the
CR bound in any observational regime\footnote{However, in
  Section~\ref{rm_ideal_low_snr} we demonstrate that in the low SNR
  limit the LS estimator can asymptotically approach the CR
  bound.}. The log-likelihood function associated to
equation~(\ref{eq_pre_5}) in this case is given by:

\begin{equation} \label{log-like}
\ln L(I^n; x_c) = {\sum_{i=1}^n \left( I_i\cdot \ln \lambda_i(x_c) -
    \lambda_i(x_c) - \ln I_i!  \right) },
\end{equation}
and we have the following result:

\begin{preposition}\label{pro_achie_astrometry}
	For any fixed and unknown parameter $x_c\in \mathbb{R}$, and
        any unbiased estimator $\tau_n$
	\begin{equation}\label{eq_main_sec_1}
		Var(\tau_n(I^n)) > \sigma_{CR}^2, 
	\end{equation}
	where $I^n$ follows a Poisson pmf ($f_{\lambda(x_c)}
          \equiv f_{x_c}$ hereafter, to shorten notation) from
          equation~(\ref{eq_pre_5}). (The proof is presented in
        Appendix \ref{proof_pro_achie_astrometry}).
\end{preposition}

The non-achievability condition imposed by
  Proposition~\ref{pro_achie_astrometry} supports the adoption of {\it
    alternative criteria} for position estimation, being ML and the
  classical LS two of the most commonly adopted approaches. The ML
  estimate of the position $\tau_{ML}(I^n)$ is obtained through the
  following rule:

\begin{equation}\label{ml1}
\tau_{ML}(I^n) = \arg \max_{\alpha\in \mathbb{R}} \ln L(I^n;\alpha),
\end{equation}
where ``$\arg \max$'' represents the argument that maximizes the
expression, while $\alpha$ is a generic variable representing the
astrometric position. Imposing the first order condition on this
optimization problem, it reduces to satisfying the condition $\frac{d
  \ln L(I^n;x_c)}{d x_c} = 0$, and, consequently, we can work with the
general expression given by
equation~(\ref{eq_proof_pro_achie_astrometry_2}). We note that a
well-known statistical result indicates that in the case of
independent and identically distributed samples the ML approach is
asymptotically unbiased and efficient (i.e., it achieves the CR bound
when the number of observations goes to infinity \citep[Chapter
  7.5]{kay93}). However for the (still independent) but
non-identically distributed setting of astrometry described by
equation~(\ref{eq_pre_5}), this asymptotic result, to the best of our
knowledge, has not been proven, and remains an open problem.

On the other hand, a version of the LS estimator (given the model
presented in Section \ref{sub_sec_astro_photo}) corresponds to the
solution of:

\begin{equation}\label{eq_subsec_mse_of_LS_1}
		\tau_{LS}(I^n)= \arg \min_{\alpha\in \mathbb{R}}
                \underbrace{\sum_{i=1}^{n} \left( I_i
                  -\lambda_i(\alpha) \right)^2}_{\equiv
                  J(I^n,\alpha)},
\end{equation}
with $\lambda_i(\alpha)=\tilde{F} g_i(\alpha)+\tilde{B}_i$ and where
$g_i(\cdot)$ is given by equation~(\ref{eq_pre_3})\footnote{Note
  that if we had assumed a detection process subject to a purely
  Gaussian noise with an rms of $\sigma$, then the probability mass
  function for each individual observation would have been given by
  $f_{\lambda}(x)=\frac{1}{\sqrt{2\pi} \sigma}
  e^-\frac{(x-\lambda)^2}{2 \sigma^2}$. In this case the
  log-likelihood function (i.e., the equivalent of
  equation~(\ref{log-like})) would be given by $\ln L(I^n; x_c) = -n
  \ln( \sqrt{2\pi} \sigma )- \frac{1}{2 \sigma^2}\sum_{i=1}^n \left(
  I_i -\lambda_i(x_c) \right)^2$. Therefore, in this scenario, finding
  the maximum of the log-likelihood would be the same as finding the
  minimum of the LS, as in
  equation~(\ref{eq_subsec_mse_of_LS_1}). This is a well-established
  result, described in many statistical books.}.

In a previous paper \citep[Section 5]{2013mendez}, we have carried out
numerical simulations using equations~(\ref{ml1})
and~(\ref{eq_subsec_mse_of_LS_1}), and have demonstrated that both
approaches are reasonable. However, an inspection of \citet[Table
  3]{2013mendez} suggests that the LS method exhibits a loss of
optimality at high-SNR in comparison with either the (Poisson
variance-) weighted LS or the ML method. This motivates a deeper study
of the LS method, to properly understand its behavior and limitations
in terms of its MSE and possible statistical bias, which is the focus
of the following Section.
%

\subsection{Bounding the performance of the Least-Squares estimator} \label{subsec_mse_of_LS}

The solution to equation~(\ref{eq_subsec_mse_of_LS_1}) is non-linear
(see Figure~\ref{fig_behave}), and it does not have a closed-form expression.
Consequently, a number of iterative approaches have been adopted (see,
e.g., \citet{stetson1987daophot,mighell2005stellar}) to solve or
approximate $\tau_{LS}(I^n)$. Hence as $\tau_{LS}(I^n)$ is implicit,
it is not possible to compute its mean, its variance, nor its
estimation error directly. We also note that, since we will be
  mainly analyzing the behavior of $\tau_{LS}(I^n)$, none of the
  caveats concerning the properness of the likelihood function
  (equation~(\ref{eq_pre_5})) raised in
  Section~\ref{sub_sec_astro_photo} are relevant in what follows,
  except for what concerns the adequacy of equation~(\ref{eq_pre_2b}),
  which we take as a valid description of the underlying flux
  distribution.

The problem of computing the MSE of an estimator that is the solution
of an optimization problem has been recently addressed by
\citet{dsp2013} using a general framework. Their basic idea was to
provide sufficient conditions on the objective function, in our case
$J(I^n,\alpha)$, to derive a good approximation for
$\mathbb{E}_{I^n\sim f_{x_c}} \left\lbrace\left( \tau_{LS}(I^n) -x_c
\right)^2\right\rbrace$.  Based on this idea, we provide below a
refined result (specialized to our astrometry problem), which relaxes
one of the idealized assumptions proposed in \citet[their
  equation~(5)]{dsp2013}, and which is not strictly satisfied in our
problem (see Remark~\ref{remark2} in Section~\ref{anaint}).  As a
consequence, our result offers upper and lower bounds for the bias and
MSE of $\tau_{LS}(I^n)$, respectively.


\begin{theorem}\label{ls_performances_bounds}
	Let us consider a fixed and unknown parameter $x_c\in
        \mathbb{R}$, and that $I^n\sim f_{x_c}$. In addition, let us
        define the residual random variable\footnote{As a short hand
          notation: $J'(I^n,x_c)\equiv \frac{d J(I^n,\alpha)}{d
            \alpha}|_{\alpha=x_c}$, and $J''(I^n,x_c)\equiv \frac{d^2
            J(I^n,\alpha)}{d^2 \alpha}|_{\alpha=x_c}$.} $W(I^n,
        \alpha) \equiv \frac{J''(I^n,\alpha) -\mathbb{E}_{I^n\sim
            f_{x_c}} \left\{ J''(I^n,\alpha) \right\}
        }{\mathbb{E}_{I^n\sim f_{x_c}} \left\{ J''(I^n,\alpha)
          \right\} }$.  If there exists $\delta \in (0,1)$ such that
        $\mathbb{P}(W(I^n,x_c)\in (-\delta,\delta))=1$, then:
	\begin{eqnarray}
		\label{eq_subsec_mse_of_LS_2a}
		&& \left| \mathbb{E}_{I^n\sim f_{x_c}} \left\{
                \tau_{LS}(I^n) \right\} - x_c \right| \leq
                \epsilon(\delta), \\
		\label{eq_subsec_mse_of_LS_2b}
		&& \mathbb{E}_{I^n\sim f_{x_c}} \left\lbrace\left( \tau_{LS}(I^n)
                -x_c \right)^2\right\rbrace \in \left(
                \frac{\sigma^2_{LS}(n)}{(1+\delta)^2},
                \frac{\sigma^2_{LS}(n)}{(1-\delta)^2} \right),
	\end{eqnarray}
where
\begin{eqnarray}
\label{nominal}
\sigma^2_{LS}(n) \equiv \frac{\mathbb{E}_{I^n\sim f_{x_c}}
  \left\{ J'(I^n,x_c)^2 \right\}}{ \left( \mathbb{E}_{I^n\sim f_{x_c}}
  \left\{ J''(I^n,x_c) \right\} \right)^2}
\end{eqnarray}
and
\begin{eqnarray}
\label{bias}
\epsilon(\delta) \equiv
\frac{\mathbb{E}_{I^n\sim f_{x_c}} \left\lbrace\left| J'(I^n,x_c)
  \right|\right\rbrace}{\mathbb{E}_{I^n\sim f_{x_c}} \left\{ J''(I^n,x_c) \right\} } \cdot
\frac{\delta}{1-\delta}.
\end{eqnarray}

\end{theorem}
(The proof is presented in Appendix
\ref{proof_ls_performances_bounds}).

\subsection{Analysis and interpretation of Theorem \ref{ls_performances_bounds}} \label{anaint}

\begin{remark}\label{rem_1}
Theorem \ref{ls_performances_bounds} is obtained under a bounded
condition (with probability one) over the
random variable $W(I^n, x_c)$. To verify whether this condition is
actually met, it is therefore important to derive an explicit
expression for $W(I^n, x_c)$. Starting from
equation~(\ref{eq_subsec_mse_of_LS_1}) it follows that:
$$J''(I^n,\alpha) = 2\sum_{i=1}^n \left[ \left( \frac{d
    \lambda_i(\alpha)}{d \alpha} \right) ^2 + (\lambda_i(\alpha) -I_i) \cdot
  \frac{d^2 \lambda_i(\alpha)}{d^2 \alpha} \right]$$
and, consequently, $\mathbb{E}_{I^n\sim f_{x_c}} \left\{ J''(I^n,x_c)
\right\}= 2\sum_{i=1}^n \left( \frac{d \lambda_i(\alpha)}{d
  \alpha}|_{\alpha=x_c} \right) ^2$. Therefore:
\begin{equation}\label{eq_subsec_mse_of_LS_3}
W(I^n, x_c)=\sum_{i=1}^n (\lambda_i(x_c)-I_i)\cdot \left[
  \lambda_i''(x_c) /\sum_{j=1}^n (\lambda_j'(x_c))^2 \right].
\end{equation}
Then, $W(I^n, x_c)$ is not bounded almost surely, since $I_i$ could
take any value in $\mathbb{N}$ with non-zero probability.  However,
$\mathbb{E}_{I^n\sim f_{x_c}} \left\{ W(I^n, x_c) \right\} =0$, and
its variance in closed-form is:
\begin{equation}\label{eq_subsec_mse_of_LS_3b}
Var(W(I^n, x_c))=\sum_{i=1}^n \lambda_i(x_c) \cdot
\left[(\lambda_i''(x_c))^2 /\left(\sum_{j=1}^n (\lambda_j'(x_c))^2
  \right)^2\right].
\end{equation}
From this, we can evaluate how far we are from the bounded assumption
of Theorem~\ref{ls_performances_bounds}. To do this, we can resort to
{\em Markov's inequality} \citep{cover_2006}, where
$\mathbb{P}(W(I^n,x_c) \notin (-\rho,\rho)) \leq Var(W(I^n,
x_c))/\rho^2 $.  Then, for any $\epsilon \in(0,1)$, we can
characterize a critical $\delta(\epsilon)>0$ such that
$\mathbb{P}(W(I^n,x_c)\in
(-\delta(\epsilon),\delta(\epsilon)))>1-\epsilon$.  Using this result
and Theorem~\ref{ls_performances_bounds}, we can bound the conditional
bias and conditional MSE of $\tau_{LS}(I^n)$ using
equations~(\ref{eq_subsec_mse_of_LS_2a})
and~(\ref{eq_subsec_mse_of_LS_2b}), respectively.  In Section
\ref{subsec_empirical}, we conduct a numerical analysis, where it is
shown that the bounded assumption for $W(I^n,x_c)$ is indeed satisfied
for a number of important realistic experimental settings in
astrometry (with very high probability).
\end{remark}

\begin{remark}\label{remark2}
	Concerning the MSE of the LS estimator,
        equation~(\ref{eq_subsec_mse_of_LS_2b}) offers a lower and
        upper bound in terms of a {\em nominal value}
        $\sigma^2_{LS}(n)$ (given by equation~(\ref{nominal})), and an
        interval around it.
In the interesting regime where $\delta\ll1$ (this regime approaches
the ideal case $\delta=0$ studied by \citet{dsp2013} in which case the
variable $W(I^n,x_c)$ becomes deterministic), we have that
$\tau_{LS}(\cdot)$ is an unbiased estimator, as shown by
equation~(\ref{eq_subsec_mse_of_LS_2a}), and, furthermore:
	$$Var(\tau_{LS}(I^n))=\mathbb{E}_{I^n\sim f_{x_c}} \left\lbrace\left(
        \tau_{LS}(I^n) -x_c \right)^2\right\rbrace=\sigma^2_{LS}(n)
        \geq \sigma_{CR}^2,$$ from
        equations~(\ref{eq_subsec_mse_of_LS_2b})
        and~(\ref{cr_scalar}).  Thus, it is interesting to provide an
        explicit expression for $\sigma^2_{LS}(n)$ which will be valid
        for the MSE of the LS method in this regime. First we note
        that, $J'(I^n,x_c)= 2 \cdot \left( \sum_{i=1}^n
        \frac{d\lambda_i(x_c)}{dx_c}\cdot(\lambda_i(x_c)-I_i) \right)$, and
        therefore:
	\begin{eqnarray}
	(J'(I^n,x_c))^2=&&4 \cdot \left(\sum_{i=1}^n\sum_{j=1,j\neq
            i}^n(I_iI_j-I_i\lambda_j-I_j\lambda_i+\lambda_i\lambda_j)
          \cdot \frac{d\lambda_i}{dx_c} \cdot
          \frac{d\lambda_j}{dx_c}\right)\nonumber\\
&&+4 \cdot \left(\sum_{i=1}^n(I_i^2-2I_i\lambda_i+\lambda_i^2) \cdot
          \left(\frac{d\lambda_i}{dx_c}\right)^2\right).\nonumber
            \end{eqnarray}
        Therefore $\mathbb{E}_{I^n\sim f_{x_c}} \left\{ J'(I^n,x_c)^2
        \right\}=4 \cdot \sum_{i=1}^n \lambda_i(x_c)\cdot
        (\lambda_i'(x_c))^2$, which implies that:
	\begin{equation}\label{eq_subsec_mse_of_LS_4}
		\sigma^2_{LS}(n) = \frac{\sum_{i=1}^n
                  \lambda_i(x_c)\cdot (\lambda_i'(x_c))^2}{\left(
                  \sum_{i=1}^n (\lambda_i'(x_c))^2\right)^2} =
                \frac{\sum_{i=1}^n (\tilde{F}g_i(x_c)+\tilde{B}_i)
                  \cdot (g_i'(x_c))^2}{\left(\tilde{F}\sum_{i=1}^n
                  (g_i'(x_c))^2\right)^2}.
	\end{equation}
	In the next section, we provide a numerical analysis to
        compare the predictions of
        equation~(\ref{eq_subsec_mse_of_LS_4}) with the CR bound
        computed through equation~(\ref{fi_astrometry}). We also
        analyze if
	this nominal value is representative of the performance of the LS estimator.
\end{remark}

\begin{remark}(Idealized Low SNR regime)
	\label{rm_ideal_low_snr}
        Following the ideal scenario where $\delta \ll 1$, we explore
        the weak signal case in which $\tilde{F}g_i(x_c) \ll
        \tilde{B}_i$ considering a constant background across the
        pixels, i.e., $\tilde{B}_i=\tilde{B}$ for all $i$.  Then
        adopting equation~(\ref{eq_subsec_mse_of_LS_4}) we have that:
	\begin{equation}\label{eq_subsec_mse_of_LS_5}
			\sigma^2_{LS}(n) \approx
                        \frac{\tilde{B}}{\tilde{F}^2\sum_{i=1}^n
                          (g_i'(x_c))^2}.
	 \end{equation}

	 On the other hand, from equation~(\ref{fi_astrometry}) we
         have that $\mathcal{I}_{x_c}(n) \approx
         {\tilde{F}^2}/{\tilde{B}}\sum_{i=1}^n (g_i'(x_c))^2$.
         Remarkably in this context,
	 the LS estimator is optimal in the sense that it approaches
         the CR bound asymptotically when a weak signal is
         observed\footnote{Recall that, according to
           Section~\ref{subsec_achie}, we have demonstrated that the
           CR can not be exactly reached in astrometry.}. This results
         is consistent with the numerical simulations in \citet[Table
           3]{2013mendez}.\footnote{We note that this asymptotic
           result can be considered the astrometric counterpart of
           what has been shown in photometry by \citet{1989perryman},
           where the LS estimator also approaches the CR bound in the
           low SNR regime.}
\end{remark}

\begin{remark}(Idealized High SNR regime)
	\label{rm_ideal_high_snr}
For the high SNR regime, assuming again that $\delta\ll 1$, we
consider the case where $\tilde{F}g_i(x_c) \gg \tilde{B}_i$ for all
$i$.  In this case:
	\begin{equation}\label{eq_subsec_mse_of_LS_6}
			\sigma^2_{LS}(n) \approx \left[ \tilde{F}
                          \frac{\left(\sum_{i=1}^n g_i'(x_c)^2
                            \right)^2}{\sum_{i=1}^n g_i(x_c)
                            g_i'(x_c)^2} \right]^{-1}\mbox{ and }
                        \sigma_{CR}^2 \approx \left[
                          \tilde{F} \sum_{i=1}^n
                          (g_i'(x_c))^2/g_i(x_c) \right]^{-1}.
	 \end{equation}
	Therefore, in this strong signal scenario, there is no match
        between the variance of the LS estimator and the CR bound, and
        consequently, we have that $\sigma^2_{LS}(n)> \sigma_{CR}^2$.
        To provide more insight on the nature of this performance gap,
        in the next proposition we offer a closed-form expression for
        this mismatch in the high-resolution scenario where the source
        is oversampled, and the size of the pixel is a small fraction
        of the width parameter $\sigma$ of the PSF in
        equation~(\ref{eq_pre_1}).
	\begin{preposition}\label{pro_gap_CR_LS_HSNR}
		Assuming the idealized high SNR regime, if we have a Gaussian-like PSF
                and $\Delta x/\sigma \ll 1$, then:
		\begin{equation}\label{eq_subsec_mse_of_LS_7}
			\frac{\sigma^2_{LS}(n)}{
                          \sigma_{CR}^2} \approx
                        \frac{8}{3\sqrt{3}}>1.
		\end{equation}
	\end{preposition}
\end{remark}
(The proof is presented in Appendix \ref{proof_pro_gap_CR_LS_HSNR}).

Equation~(\ref{eq_subsec_mse_of_LS_7}) shows that there is a very
significant performance gap between the CR bound and the MSE of the LS
estimator in the high SNR regime. This result should motivate the
exploration of alternative estimators that could approach more closely
the CR bound in this regime.

\section{Numerical analysis of the LS estimator} \label{subsec_empirical}

In this section we explore the implications of Theorem
\ref{ls_performances_bounds} in astrometry through the use of
simulated observations.  First, we analyze if the bounded condition
over $W(I^n, \alpha)$ adopted in Theorem~\ref{ls_performances_bounds}
is a valid assumption for the type of settings considered in
astronomical observations.  After that, we compare
how efficient is the LS estimator proposed in
equation~(\ref{eq_subsec_mse_of_LS_1}) as a function of the SNR and
pixel resolution, adopting for that purpose Theorem
\ref{ls_performances_bounds} and Proposition
\ref{pro_FI_photometry_astrometry}.

To perform our simulations, we adopt some realistic design variables
and astronomical observing conditions to model the problem
\citep{2013mendez,2014mendez}. For the PSF, various analytical and
semi-empirical forms have been proposed, see for instance the
ground-based model in \citet{king1971} and the space-based model in
\citet{bendinelli1987}. For our analysis, we adopt in
equation~(\ref{eq_pre_3}) the Gaussian PSF where $\phi(x,\sigma)=
\frac{1}{\sqrt{2\pi} \, \sigma} e^{-
  \frac{1}{2}\cdot\left(\frac{x}{\sigma}\right)^2}$, and where
$\sigma$ is the width of the PSF, assumed to be known. This PSF has
been found to be a good representation for typical astrometric-quality
ground-based data \citep{2010mendez}. In terms of nomenclature, the
$FWHM\equiv 2 \sqrt{2 \ln 2} \,\, \sigma$, measured in arcsec, denotes
the {\em Full-Width at Half-Maximum} ($FWHM$) parameter, which is an
overall indicator of the image quality at the observing site
\citep{chromey2010measure}\footnote{In our simulations we fix the
  value of $FWHM$ to a constant value. However, in many cases the
  image quality changes as a function of position in the field-of-view
  due to optical distortions, specially important in large focal plane
  arrays. For example, in the case of the SDSS (which consists of
  30~CCDs of 2048$^2$ pix$^2$ each, covering 2\arcdeg.3 on the sky at
  a resolution of 0.396~arcsec/pix), the $FWHM$ may vary up to 15\%
  from center to corner of {\it one} detector
  (\citet[Section~4.1]{luptonetal2001}). The impact on the CR bound of
  these changes has been discussed in some detail by
  \citet[Section~3.4]{2014mendez}.}.

The background profile, represented by $ \left\{\tilde{B}_i,
i=1,..,n\right\}$, is a function of several variables, like the
wavelength of the observations, the moon phase (which contributes
significantly to the diffuse sky background), the quality of the
observing site, and the specifications of the instrument itself.  We
will consider a uniform background across pixels underneath the PSF,
i.e., $\tilde{B_i}=\tilde{B}$ for all $i$. To characterize the
magnitude of $\tilde{B}$, it is important to first mention that the
detector does not measure photon counts (or, actually, photo-$e^-$)
directly, but a discrete variable in ``{\em Analog to Digital Units}
(ADUs)'' of the instrument, which is a linear proportion of the photon
counts \citep{gilligand1992}.  This linear proportion is characterized
by the gain of the instrument $G$ in units of $e^-$/ADU. $G$ is just a
scaling factor, where we can define $F \equiv \tilde{F}/G$ and $B
\equiv \tilde{B}/G$ as the brightness of the object and background,
respectively, in the specific ADUs of the instrument.  Then, the
background (in {ADUs}) depends on the pixel size $\Delta x$ as follows
\citep{2013mendez}:
\begin{equation}\label{eq_subsec_empirical_1}
B=f_s \cdot \Delta x+ \frac{D+RON^2}{G} \;\; \mbox{ADU},
\end{equation}
where $f_s$ is the (diffuse) sky background in ADU/arcsec (if $\Delta
x$ is measured in arcsec), while $D$ and $RON$, both measured in
$e^{-}$, model the dark-current and read-out-noise of the detector on
each pixel, respectively. Note that the first component in
equation~(\ref{eq_subsec_empirical_1}) is attributed to the site, and
its effect is proportional to the pixel size.  On the other hand, the
second component is attributed to errors of the integrating device
(detector), and it is pixel-size independent. This distinction is
central when analyzing the performance as a function of the pixel
resolution of the array (see details in
\citet[Sec. 4]{2013mendez}). More important is the fact that in
typical ground-based astronomical observation, long exposure times are
considered, which implies that the background is dominated by diffuse
light coming from the sky (the first term in the RHS of
equation~(\ref{eq_subsec_empirical_1})), and not from the detector
\citep[Sec. 4]{2013mendez}.

For the experimental conditions, we consider the scenario of a
ground-based station located at a good site with clear atmospheric
conditions and the specifications of current science-grade CCDs, where
$f_s=1502.5$~ADU/arcsec, $D=0$~$e^{-}$, $RON=5$~$e^{-}$,
$FWHM=1$~arcsec and $G=2$~$e^-$/ADU (with these values $B=313$~ADU for
$\Delta x=0.2$~arcsec using
equation~(\ref{eq_subsec_empirical_1})). In terms of scenarios of
analysis, we explore different pixel resolutions for the CCD array
$\Delta x \in [0.1, 0.7]$ measured in arcsec. Note that a change in
$\Delta x$ will be reflected upon the limits of the integral to
compute the pixel response function $g_i(x_c)$ (see
equation~(\ref{eq_pre_3})) as well as in the calculation of the
background level per pixel $\tilde{B}_i$, according to
equation~(\ref{eq_subsec_empirical_1}). Therefore, a change in $\Delta
x$ is not only a design feature of the detector device, but it implies
also a change in the distribution of the background underneath the
PSF. The impact of this covariant device-and-atmosphere change in the
CR bound is explained in detail in \citet[Section 4, see also their
  Figure~2]{2013mendez}.

In our simulations, we also consider different signal strengths
$\tilde{F} \in \left\{1\,080, 3\,224, 20\,004, 60\,160
\right\}$\footnote{These are the same values explored in \citet[Table
    3]{2013mendez}.}, measured in photo-$e^-$, corresponding to
SNR~$\in \sim \left\{12, 33, 120, 230 \right\}$
respectively\footnote{For a given $\tilde{F}$ and $f_s$ there is a
  weak dependency of SNR on the pixel size $\Delta x$, see
  equation~(28) in \citet{2013mendez}.}. Note that increasing
$\tilde{F}$ implies increasing the SNR of the problem, which can be
approximately measured by the ratio $\tilde{F}/\tilde{B} \equiv
F/B$\footnote{We note that while the ratio $\tilde{F}/\tilde{B}$ can
    be used as a proxy for SNR, in what follows we have used the exact
    expression to compute this quantity, as given by equation~(28) in
    \citet{2013mendez}.}. On a given detector plus telescope setting,
these different SNR scenarios can be obtained by changing
appropriately the exposure time (open shutter) that generates the
image.

\subsection{Analyzing the bounded condition over $W(I^n, \alpha)$}\label{anal_bound}
\begin{figure}[h]
\centering
\includegraphics[width=0.7\textwidth]{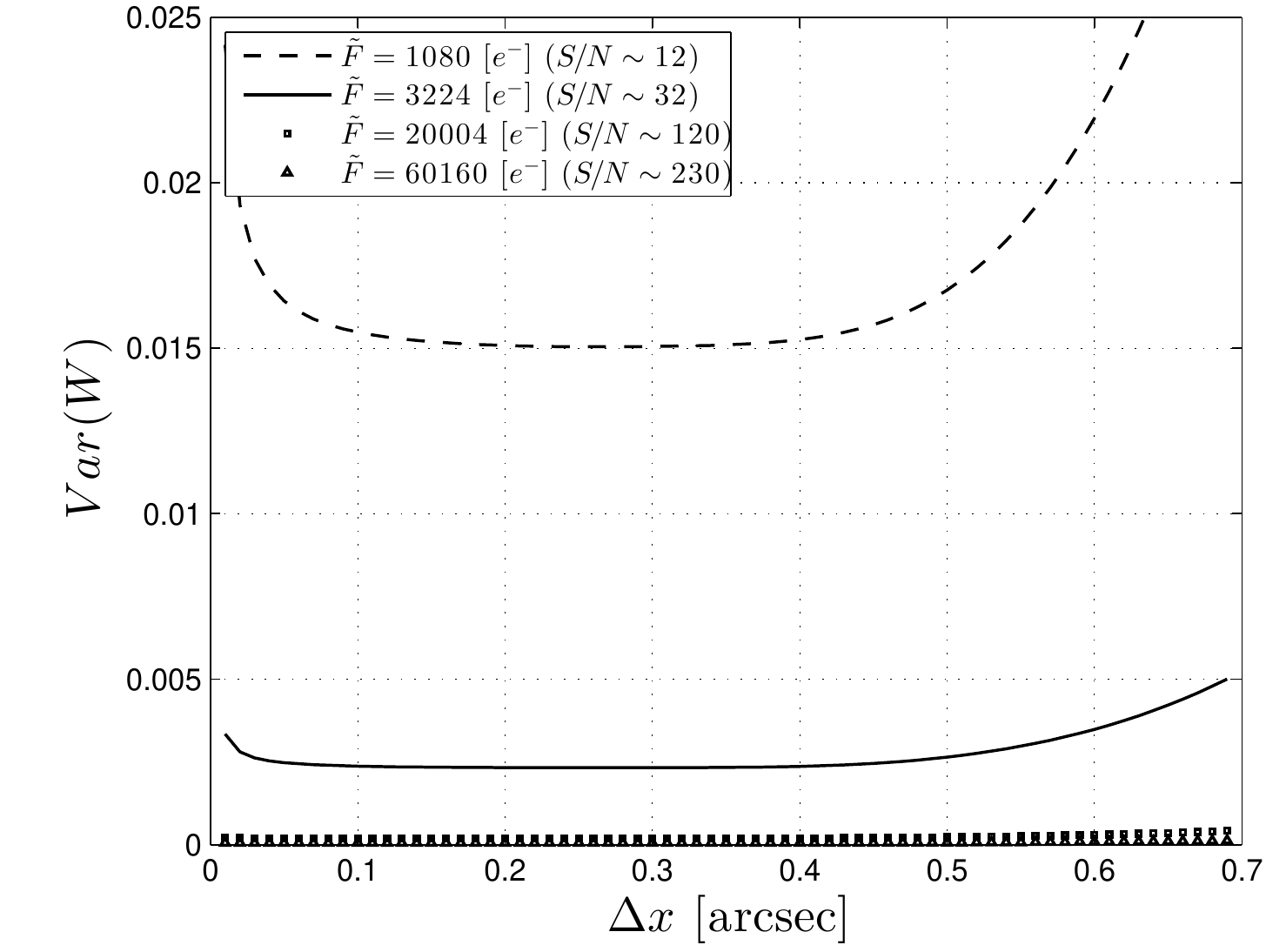}
\caption{Variance of the residual random variable $W(I^n, x_c)$
  (dimensionless) as given by equation~(\ref{eq_subsec_mse_of_LS_3b}),
  as a function of the pixel resolution $\Delta x$~arcsec for
  different realistic SNR scenarios (function of $\tilde{F}$)
  encountered in ground-based astronomical observations. Since the
  admissible range for $\delta$ is the interval (0,1), the small
  computed values indicates that the bounded assumption in Theorem~1
  can be considered as valid under these conditions.}
\label{fig1}
\end{figure}

To validate how realistic is the bounded assumption over $W(I^n,
\alpha)$ in our problem, we first evaluate the variance of $W(I^n,
x_c)$ from equation~(\ref{eq_subsec_mse_of_LS_3b}), this is presented
in Figure~\ref{fig1} for different SNR regimes and pixel resolutions
in the array. Overall, the magnitudes are very small considering the
admissible range $(0,1)$ for $W(I^n, x_c)$ stipulated in Theorem
\ref{ls_performances_bounds}. Also, given that $W(I^n, x_c)$ has zero
mean, the bounded condition will happen with high
probability. Complementing this, Figure~\ref{fig2} presents the
critical $\delta$ across different pixel resolutions and SNR
regimes\footnote{These values were computed empirically (from
  frequency counts) using $5\,000$ realizations of the random variable
  $W(I^n, x_c)$ for the different SNR regimes and pixel
  resolutions.}. For this, we fix a small value of $\epsilon$
($=10^{-3}$ in this case), and calculate $\delta$ such that
$W(I^n,x_c)\in (-\delta,\delta)$ with probability $1-\epsilon$.  From
the curves obtained, we can say that the bounded assumption is holding
(values of $\delta$ in $(0,1)$) for a wide range of representative
experimental conditions and, consequently, we can use Theorem
\ref{ls_performances_bounds} to provide a range on the performance of
the LS estimator. Note that the idealized condition of $\delta \approx
0$ is realized only for the very high SNR regime (strong signals).

\subsection{Performance Analysis of the LS estimator}
\begin{figure}[h]
\centering
\includegraphics[width=0.7\textwidth]{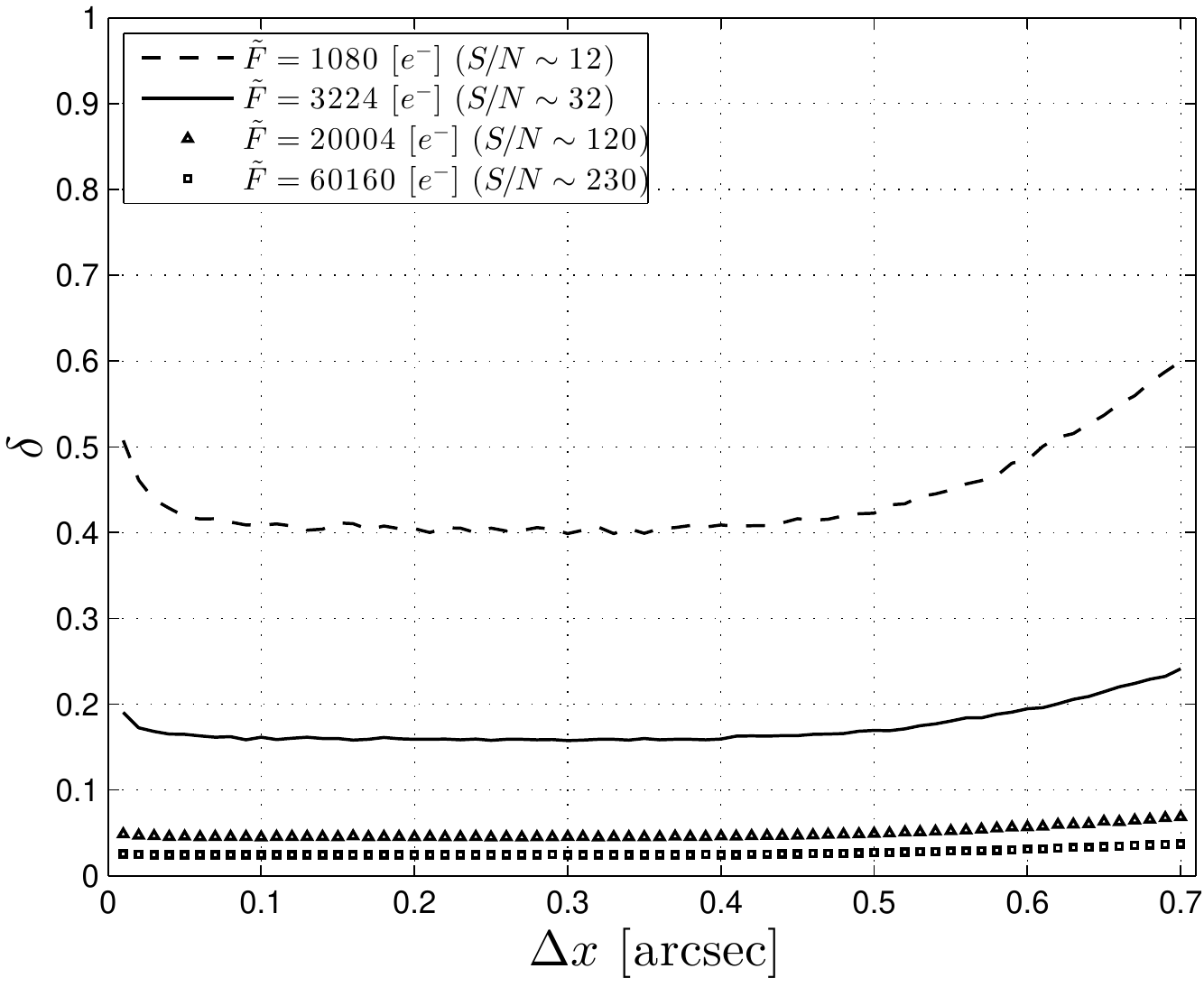}
\caption{Numerical computation of the critical $\delta$
  (dimensionless) such that $P(W(I^n,x_c)\in
  (-\delta,\delta))>1-10^{-3}$.  In all the scenarios (SNR, and
  $\Delta x$), $5\,000$ realizations of the random variable $W(I^n,
  x_c)$ are used to estimate the probability distribution for
  $W(I^n,x_c)$, and $\delta$, from frequency counts. As $\delta$
  decreases, we have a smaller bias (see
  equations~(\ref{eq_subsec_mse_of_LS_2a}) and~(\ref{bias})) and a
  narrower range for the MSE of the LS estimator
  (equation~(\ref{eq_subsec_mse_of_LS_2b})).}
\label{fig2}
\end{figure}

\begin{figure}
\plottwo{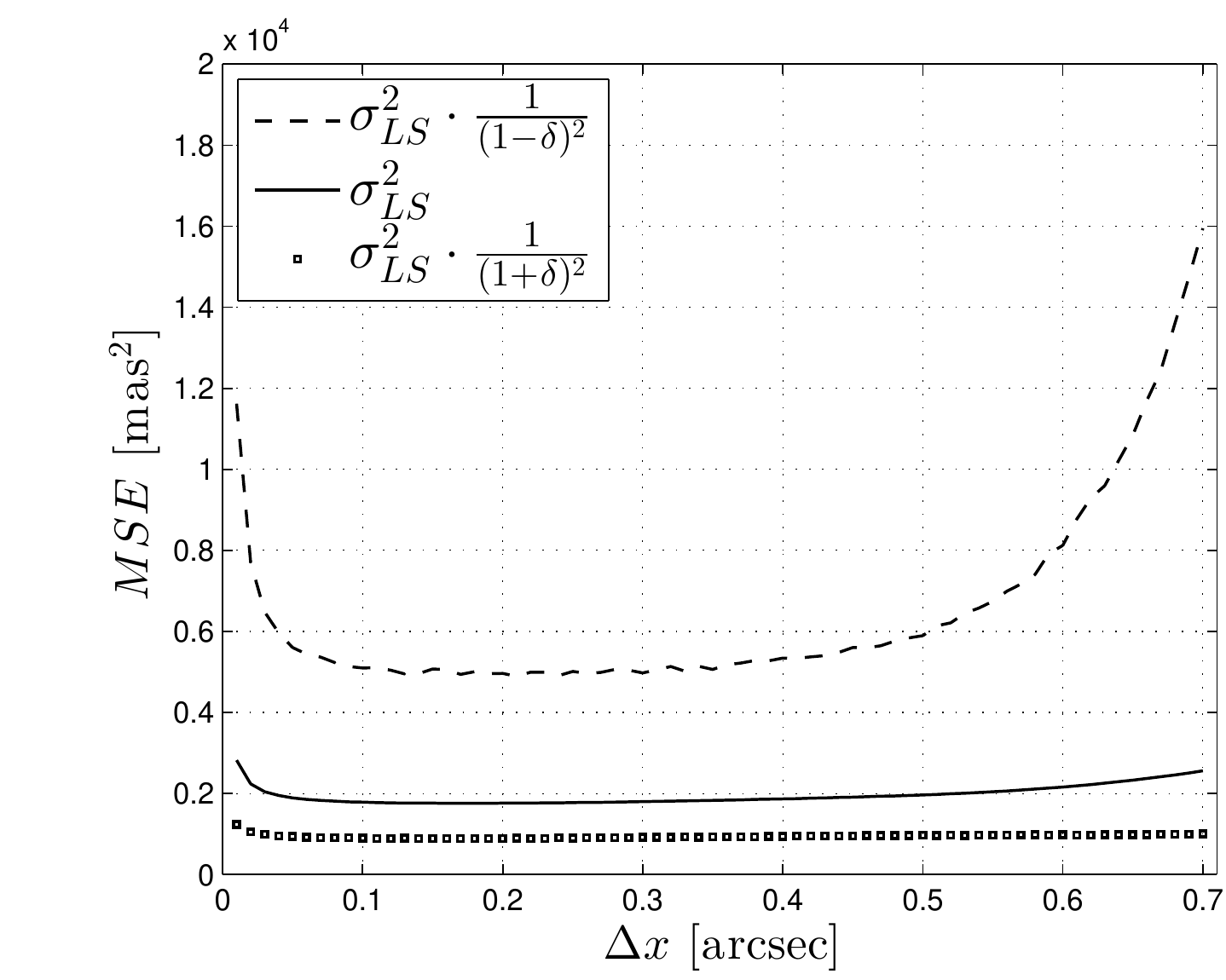}{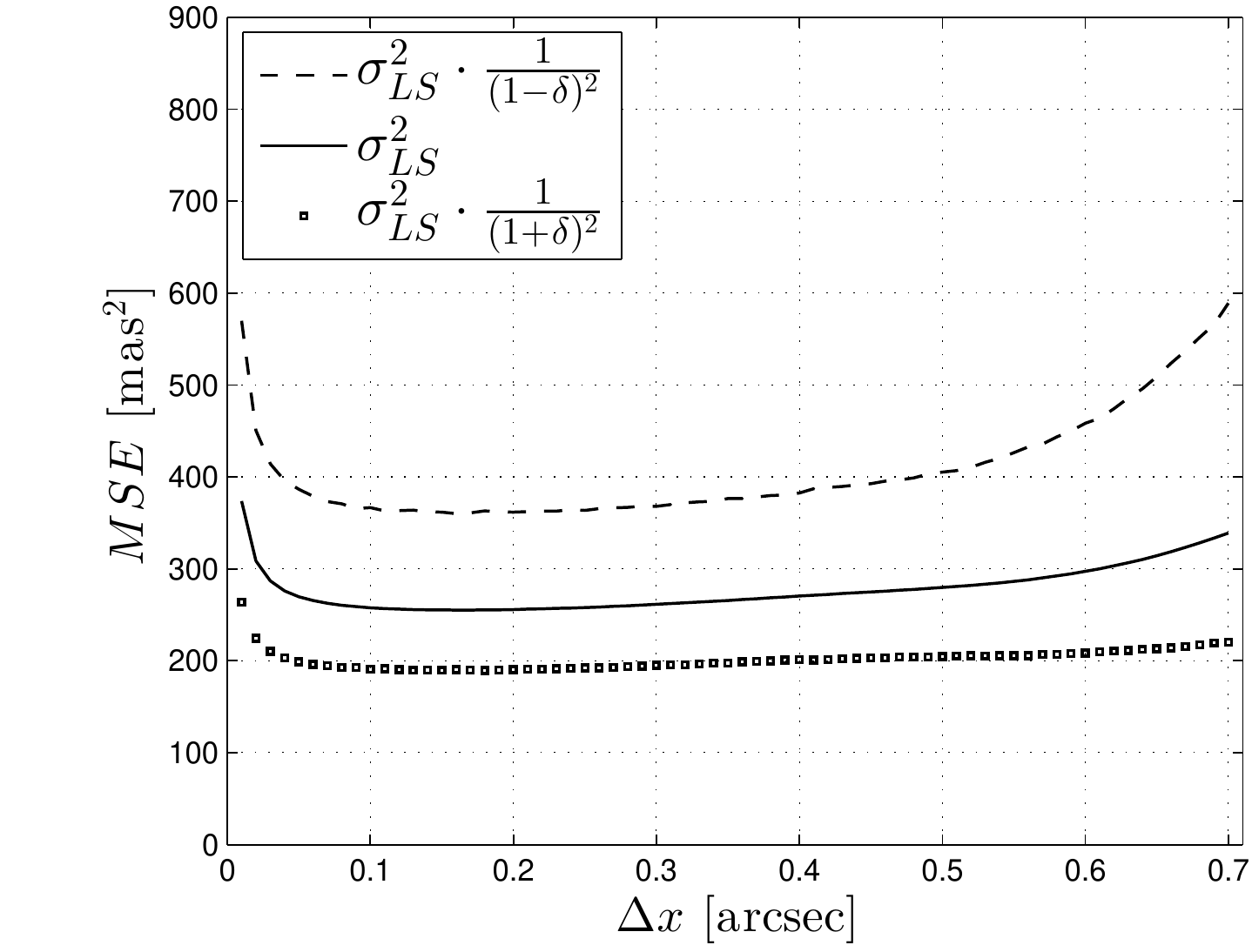}
\plottwo{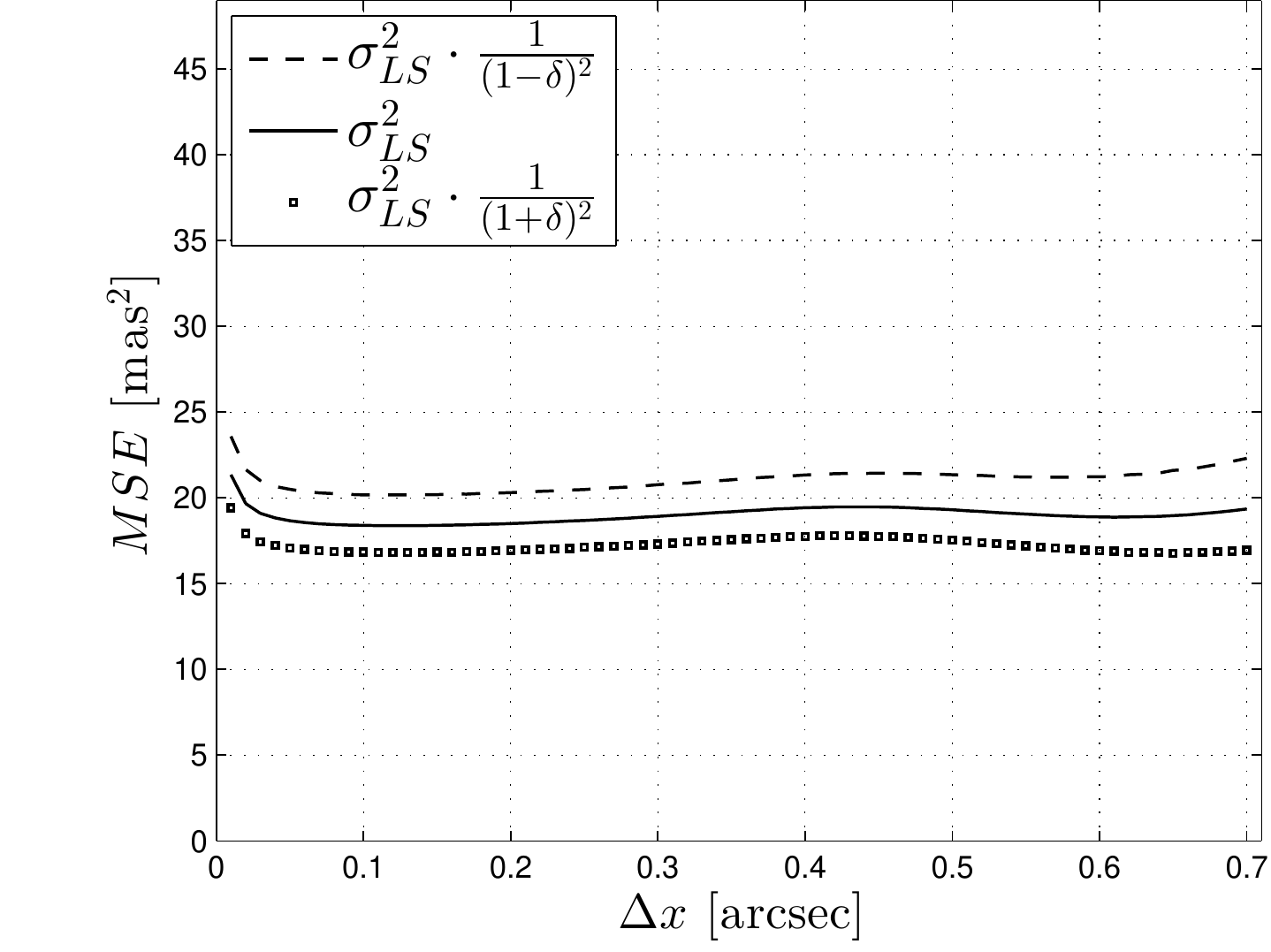}{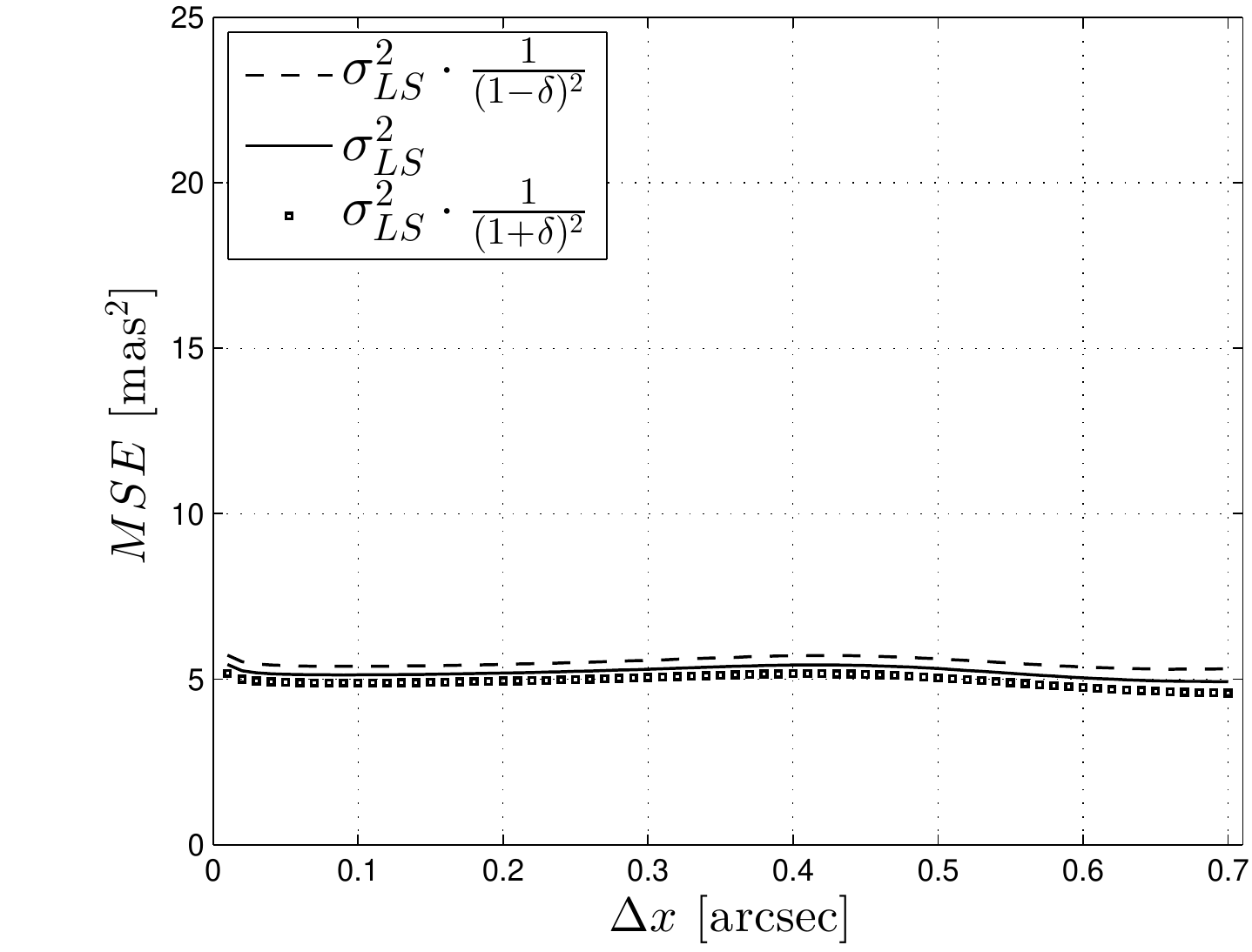}
\caption{Range of the MSE performance (in milli-arcsec$^2$ = mas$^2$)
  for the LS method in astrometry predicted by
  Theorem~\ref{ls_performances_bounds}
  (equations~(\ref{eq_subsec_mse_of_LS_2b}) and
  (\ref{nominal})). Results are reported for different representative
  values of $\tilde{F}$ and across different pixel sizes: (From
  top-left to bottom-right) $\tilde{F}= 1\,080$~e$^-$; $\tilde{F}=
  3\,224$~e$^-$; $\tilde{F}=20\,004$~e$^-$; $\tilde{F}=
  60\,160$~e$^-$.}\label{fig3}
\end{figure}

We adopt equation~(\ref{eq_subsec_mse_of_LS_2b}) which provides an
admissible range for the MSE performance of the LS estimator. For that
we use the critical $\delta$ obtained in Figure~\ref{fig2}. These
curves for the different SNR regimes and pixel resolutions are shown
in Figure~\ref{fig3}. Following the trend reported in
Figure~\ref{fig2}, the nominal value $\sigma_{LS}^2$ is a precise
indicator for the LS estimator performance for strong signals
(matching the idealized conditions stated in Remark
\ref{rm_ideal_high_snr}), while on the other hand, Theorem
\ref{ls_performances_bounds} does not indicate whether $\sigma_{LS}^2$
is accurate or not for low SNR, as we deviate from the idealized case
elaborated in Remark \ref{rm_ideal_low_snr}.  Nevertheless, we will
see, based on some complementary empirical results reported in what
follows, that even for low SNR, the nominal $\sigma_{LS}^2$ predicts
the performance of the LS estimator quite well.

\begin{figure}[h]
\centering
\includegraphics[width=0.7\textwidth]{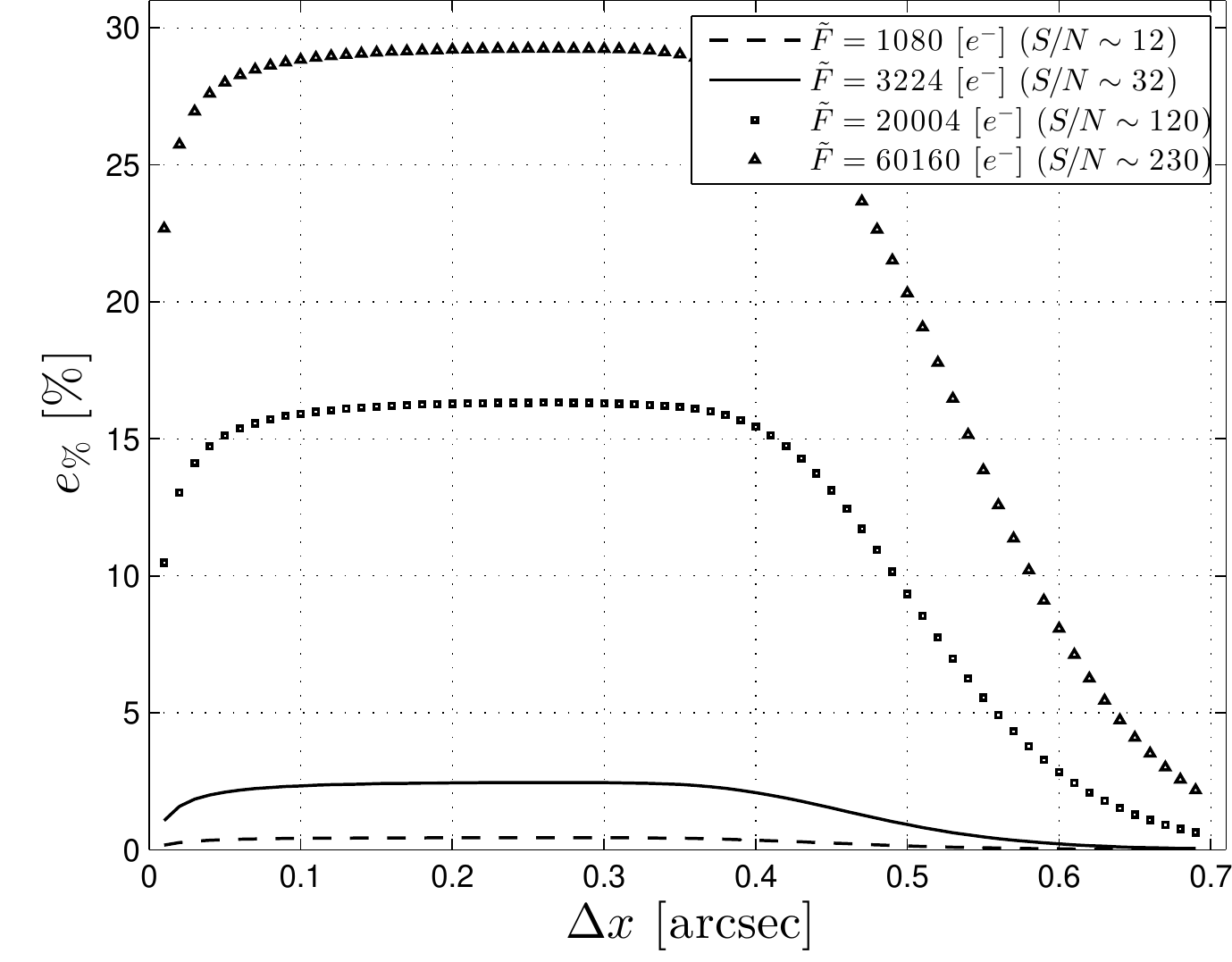}
\caption{Relative performance differences between $\sigma_{LS}^2$ in
  Theorem \ref{ls_performances_bounds} (equation~(\ref{nominal})) and
  the CR bound $\sigma_{CR}^2$ in Proposition
  \ref{pro_FI_photometry_astrometry}
  (equation~(\ref{fi_astrometry})). Results are reported for different
  SNR and pixel sizes. A significant performance gap between the LS
  technique and the CR bound is found for $FWHM$/$\Delta x < 1$ (good
  sampling of the PSF) at high SNR, indicating that, in this regime,
  the LS method is sub-optimal, in agreement with
  Proposition~\ref{pro_gap_CR_LS_HSNR} (see also
  equation~(\ref{eq_subsec_mse_of_LS_7})). This gap becomes
  monotonically smaller as the SNR decreases.}
\label{fig4}
\end{figure}

Assuming for a moment the idealized case in which $\delta\ll1$, we can
reduce the performance analysis to measuring the gap between the
nominal value predicted by Theorem \ref{ls_performances_bounds}
(equation~(\ref{nominal})), and the CR bound in Proposition
\ref{pro_FI_photometry_astrometry}. Figure~\ref{fig4} shows the
relative difference given by $e_{\%}=
\frac{\sigma_{LS}^2-\sigma_{CR}^2}{\sigma_{CR}^2} \cdot 100$.  From
the figure we can clearly see that, in the low SNR regime, the
relative performance differences tends to zero and, consequently, the
LS estimator approaches the CR bound, and it is therefore an efficient
estimator. This matches what has been stated in Remark
\ref{rm_ideal_low_snr}. On the other hand for high SNR, we observe a
performance gap that is non negligible (up to $\approx27\%$ relative
difference above the CR for $\tilde{F}=60\,160$~$e^-$, and $\approx
15\%$ above the CR for $\tilde{F}=20\,004$~$e^-$ for $\Delta
x=0.2$~arcsec). This is consistent with what has been argued in Remark
\ref{rm_ideal_high_snr}.  Note that in this regime, the idealized
scenario in which $\delta \ll 1$ is valid (see Figure~\ref{fig2}) and,
thus, $\mathbb{E}_{I^n\sim f_{x_c}} \left\lbrace\left( \tau_{LS}(I^n)
-x_c \right)^2\right\rbrace \approx \sigma_{LS}^2$, which is not
strictly the case for the low SNR regime (although see
Figure~\ref{fig7}, and the discussion that follows).

\begin{figure}
\plottwo{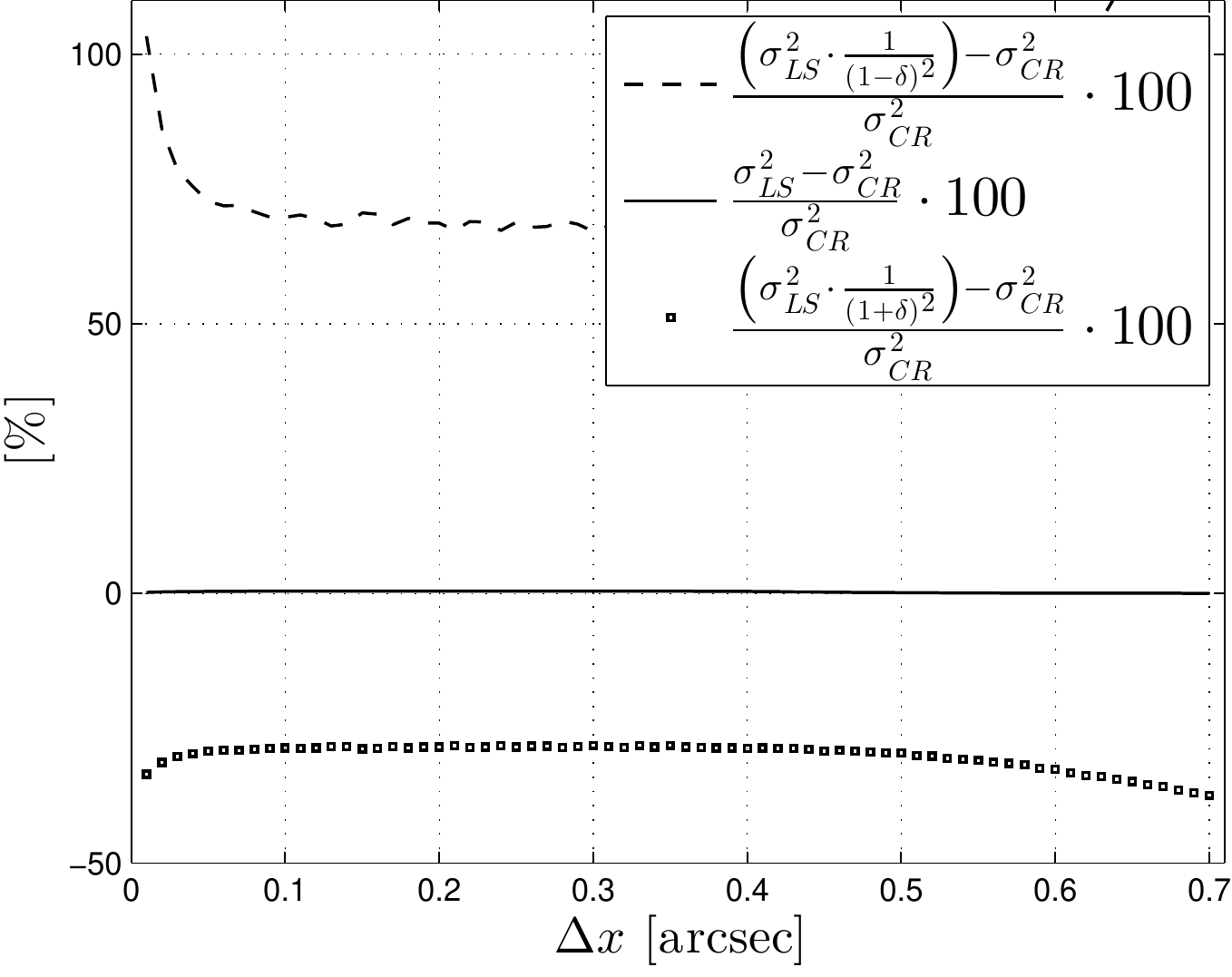}{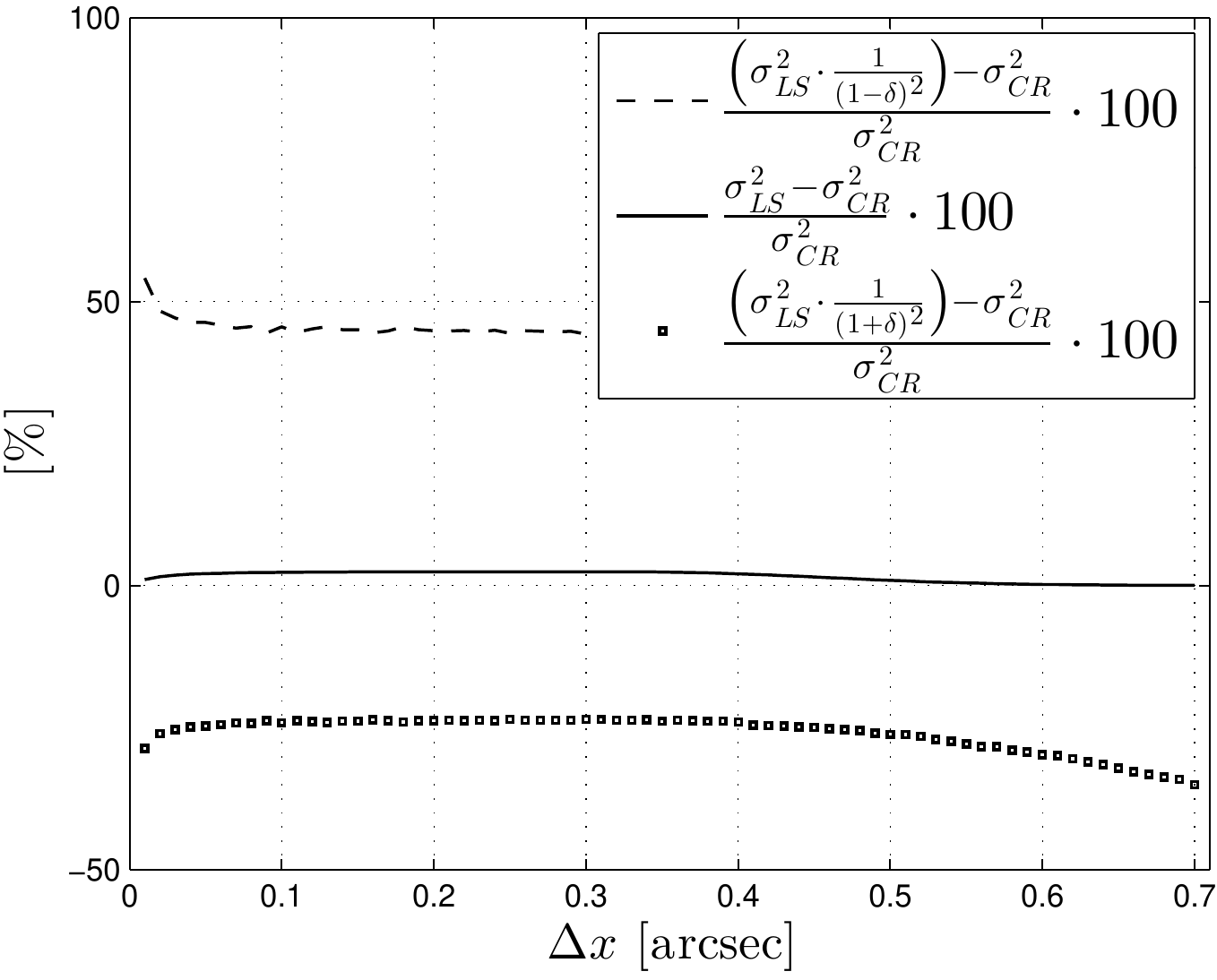}
\plottwo{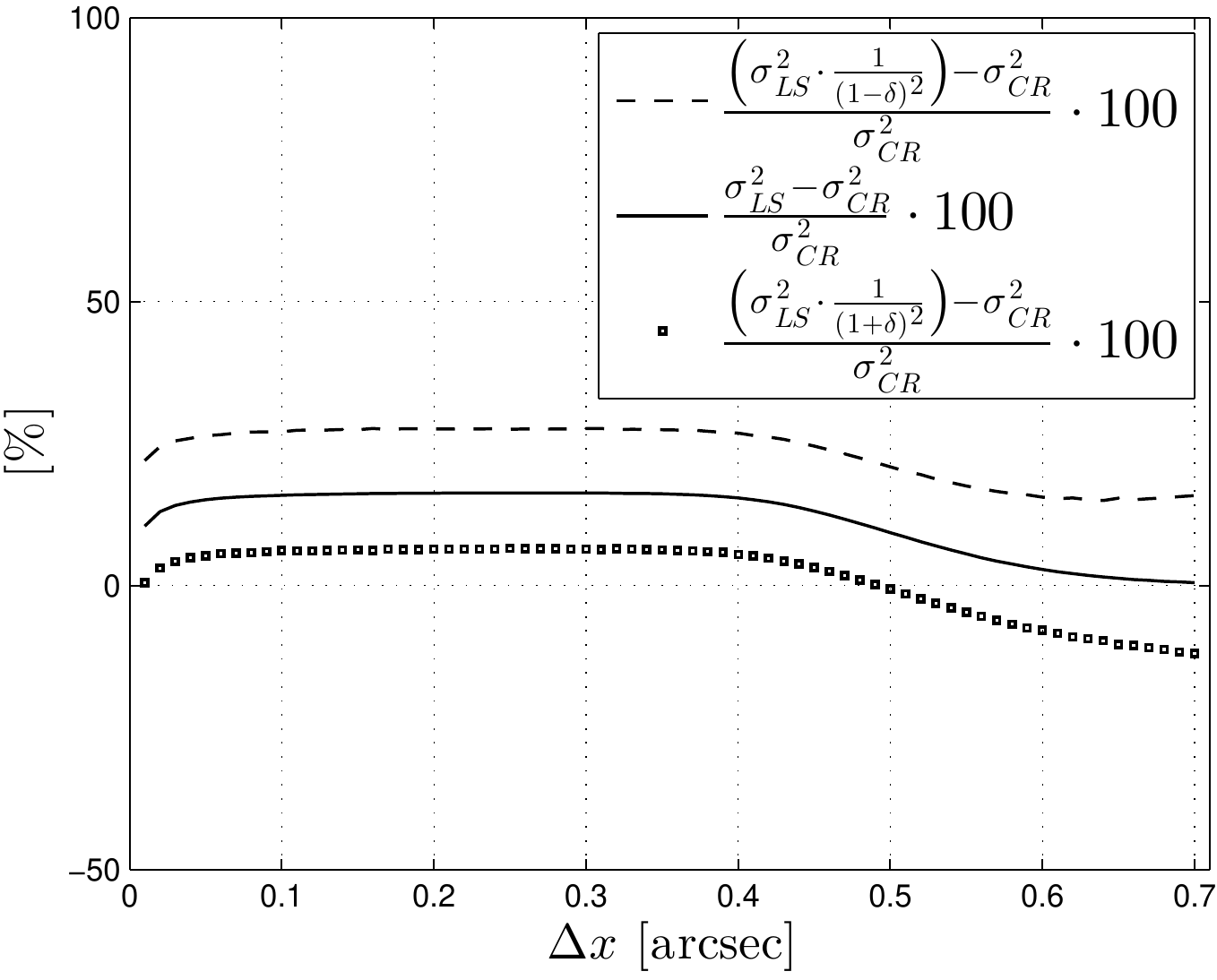}{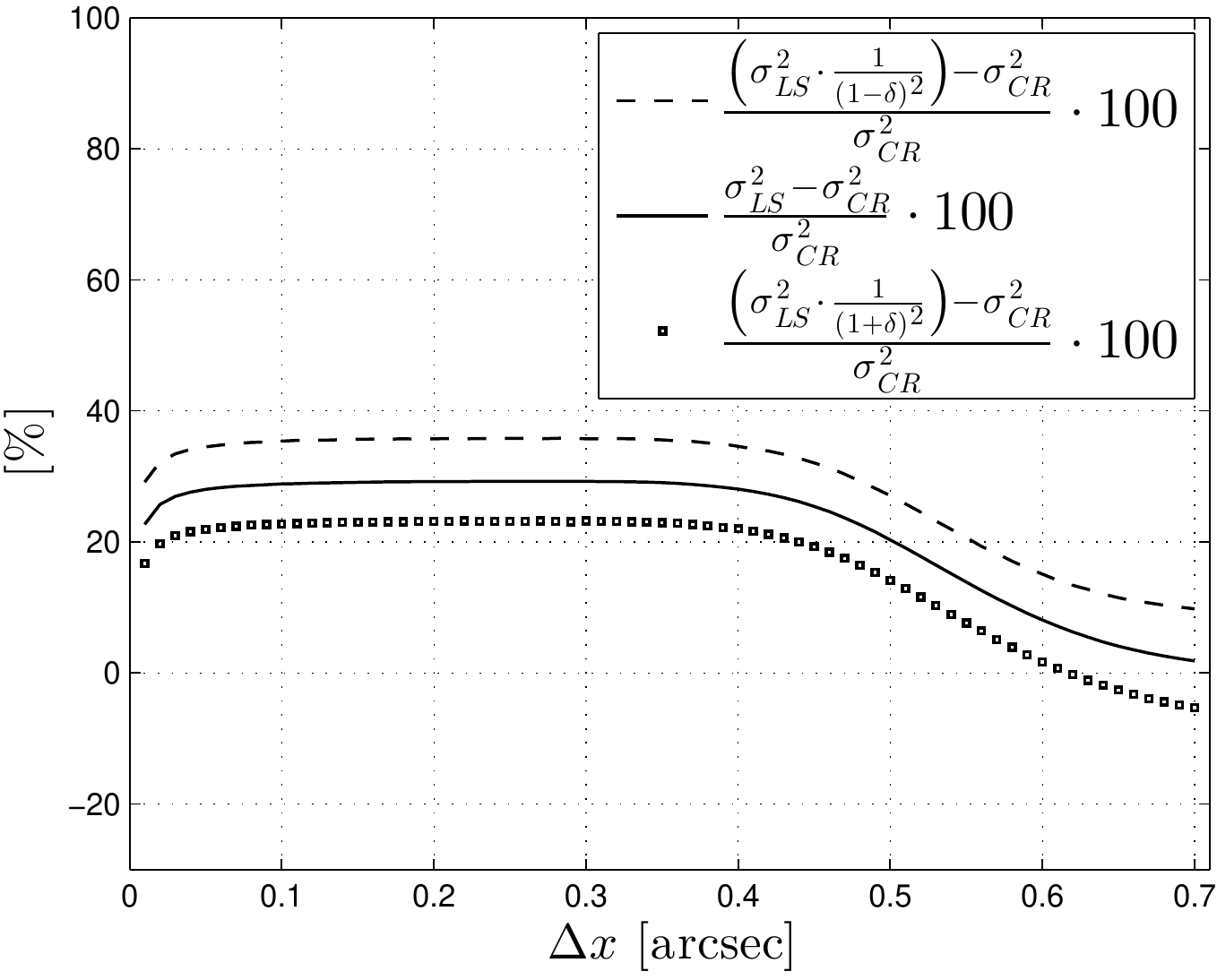}
    \caption{Relative performance differences between the range of
      performance $\left( \frac{\sigma_{LS}^2(n)}{(1+\delta)^2},
      \frac{\sigma_{LS}^2(n)}{(1-\delta)^2} \right)$ stipulated in
      Theorem~\ref{ls_performances_bounds} for the LS estimator
      (equations~(\ref{eq_subsec_mse_of_LS_2b}) and (\ref{nominal}))
      and the CR bound $\sigma_{CR}^2$ in
      Proposition~\ref{pro_FI_photometry_astrometry}
      (equation~(\ref{fi_astrometry})).  Results are reported for
      different $\tilde{F}$ and across different pixel sizes: (From
      top-left to bottom-right) $\tilde{F}=1\,080$~e$^-$;
      $\tilde{F}=3\,224$~e$^-$; $\tilde{F}=20\,004$~e$^-$;
      $\tilde{F}=60\,160$~e$^-$. The 0\% level corresponds to having
      achieved the CR bound. Note that as the SNR decreases, the bias
      (see equations~(\ref{eq_subsec_mse_of_LS_2a}) and~(\ref{bias}),
      and the possible range for the MSE of the LS method (see
      equations~(\ref{eq_subsec_mse_of_LS_2b})) increase (see also
      Figure~\ref{fig2}).}\label{fig:inter}
\end{figure}

\begin{figure}[h]
\centering
\includegraphics[width=0.7\textwidth]{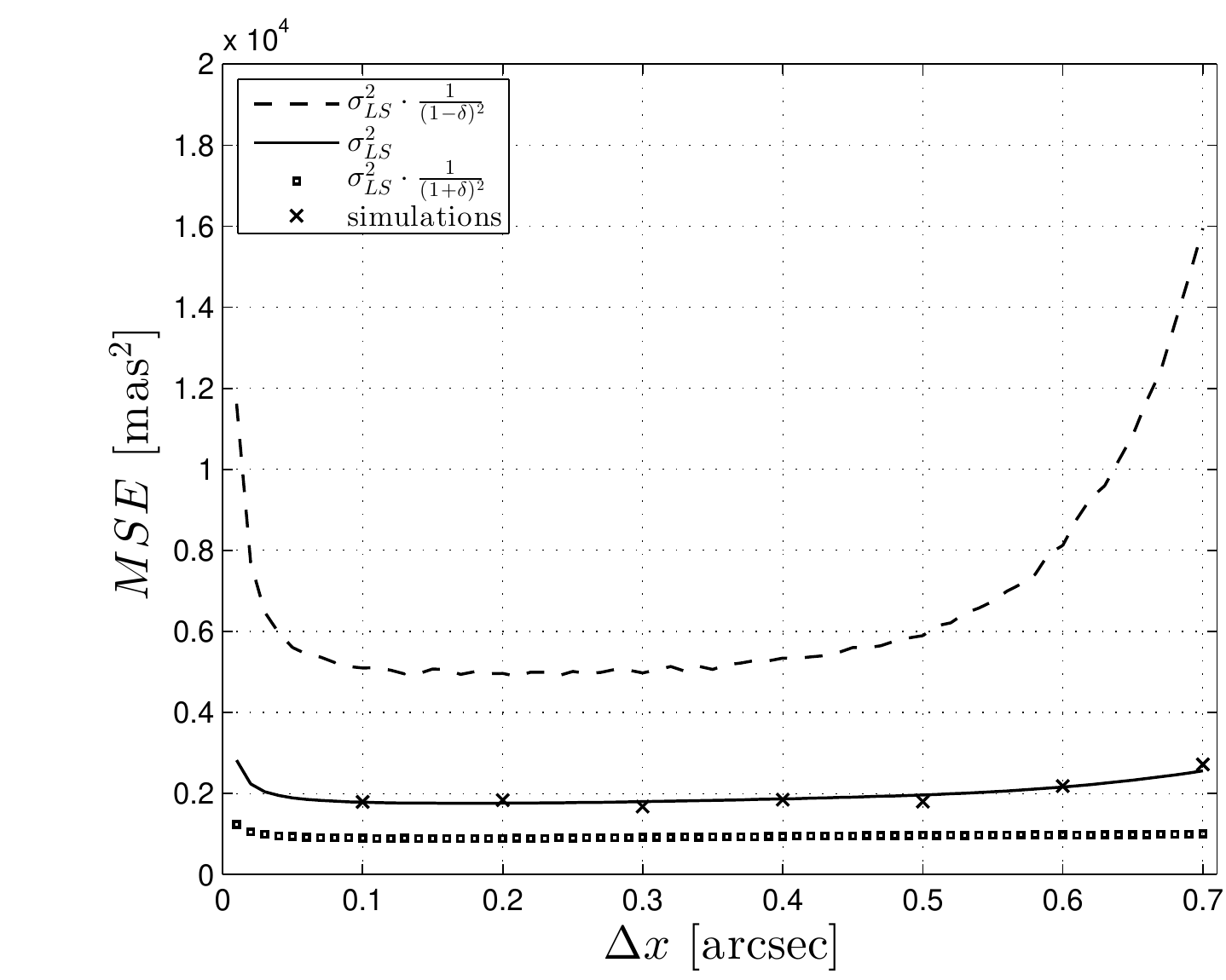}
\caption{Comparison between the nominal value $\sigma_{LS}^2$, the
  performance range $\left( \frac{\sigma_{LS}^2(n)}{(1+\delta)^2},
  \frac{\sigma_{LS}^2(n)}{(1-\delta)^2} \right)$ stipulated in Theorem
  \ref{ls_performances_bounds}, and the empirical estimation of
  $\mathbb{E}_{I^n\sim f_{x_c}} \{\left( \tau_{LS}(I^n) -x_c
  \right)^2\} $ from simulations for a low SNR regime of
  $\tilde{F}=1\,080$~e$^-$. The fact that the simulations follow
  closely the nominal value, even at low SNR, justifies the use of
  $\sigma_{LS}$ as given by equations~(\ref{nominal}) and
  (\ref{eq_subsec_mse_of_LS_4}) as a benchmark of the LS method at any
  SNR.}
\label{fig7}
\end{figure}

\begin{figure}
\plottwo{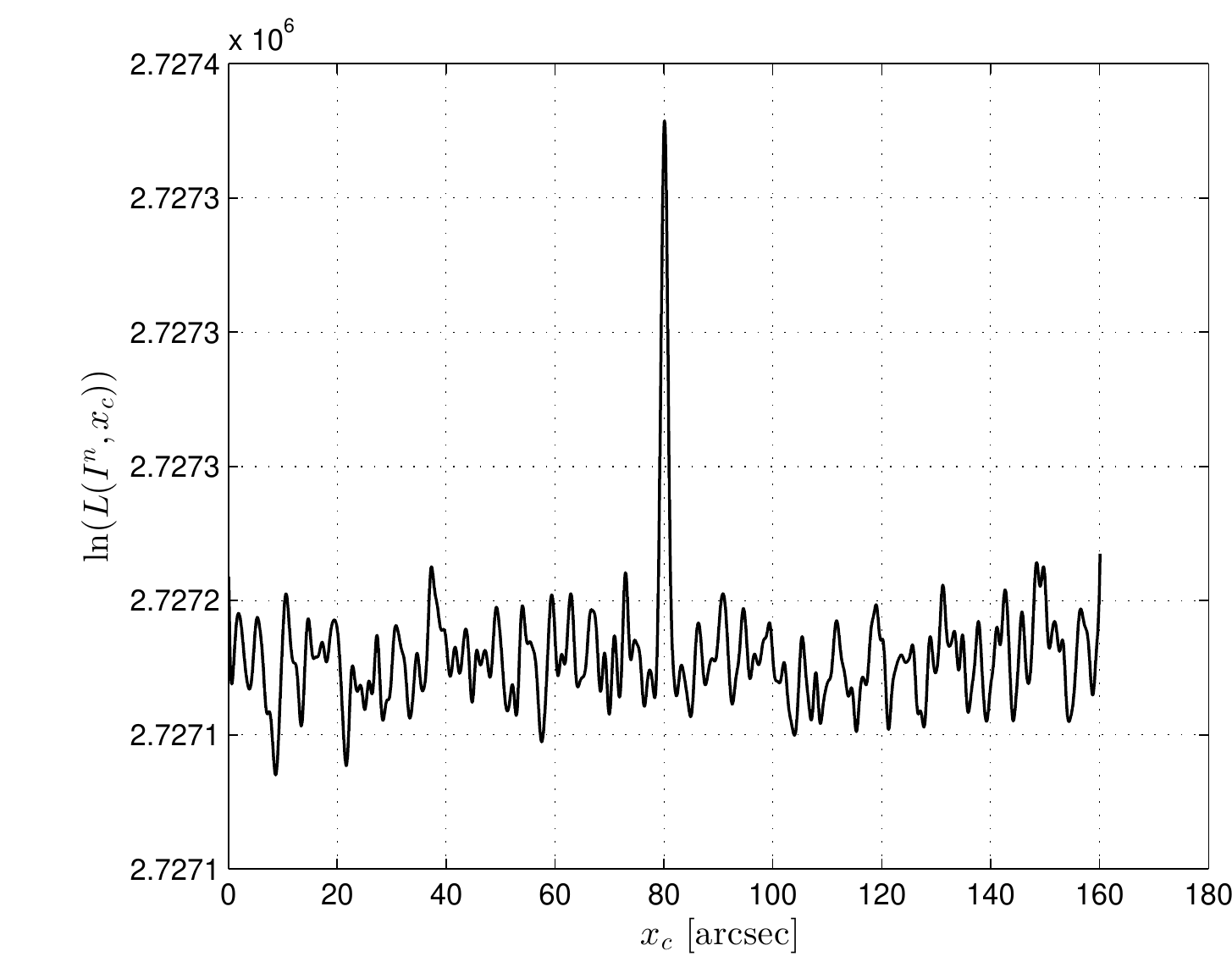}{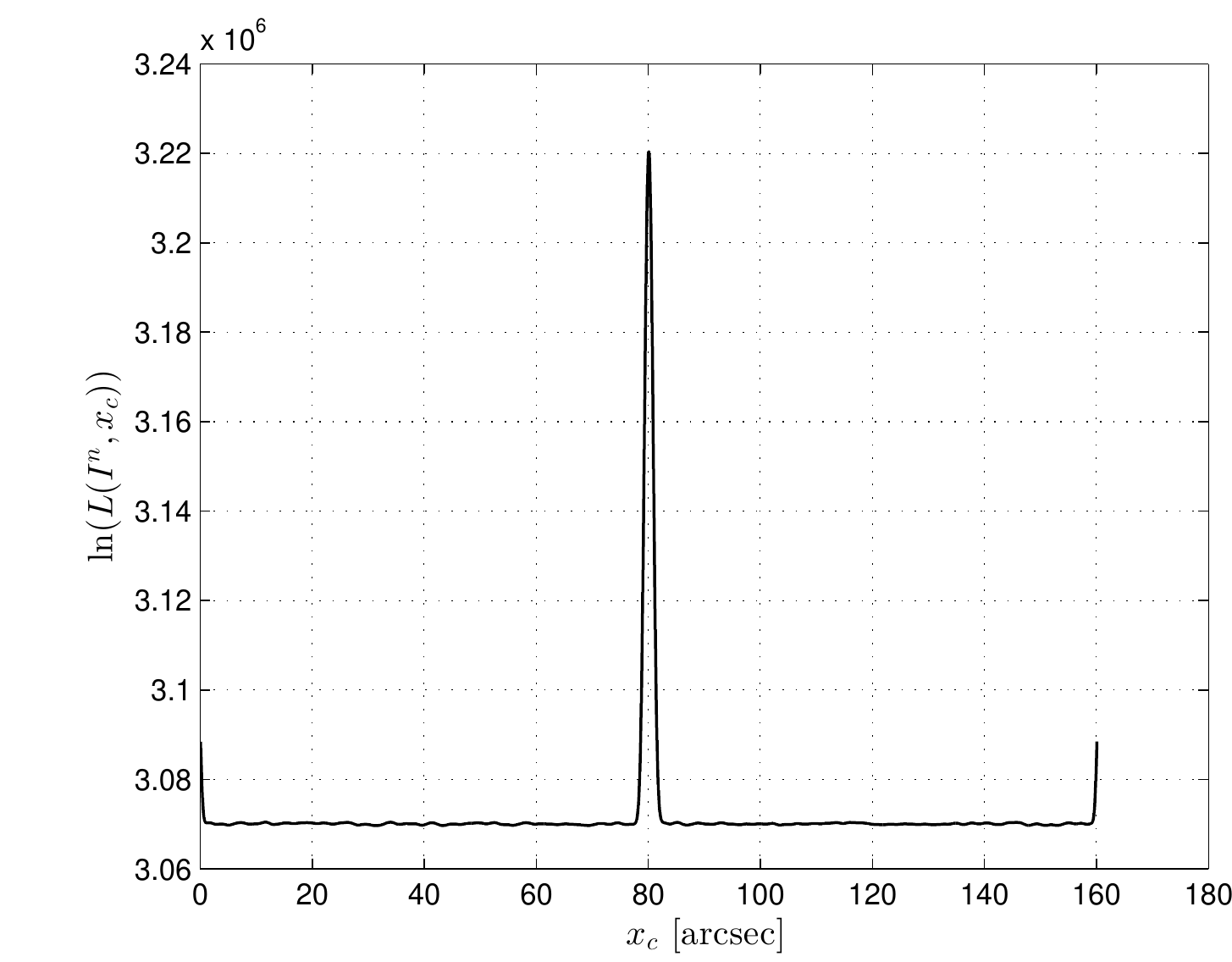}
\plottwo{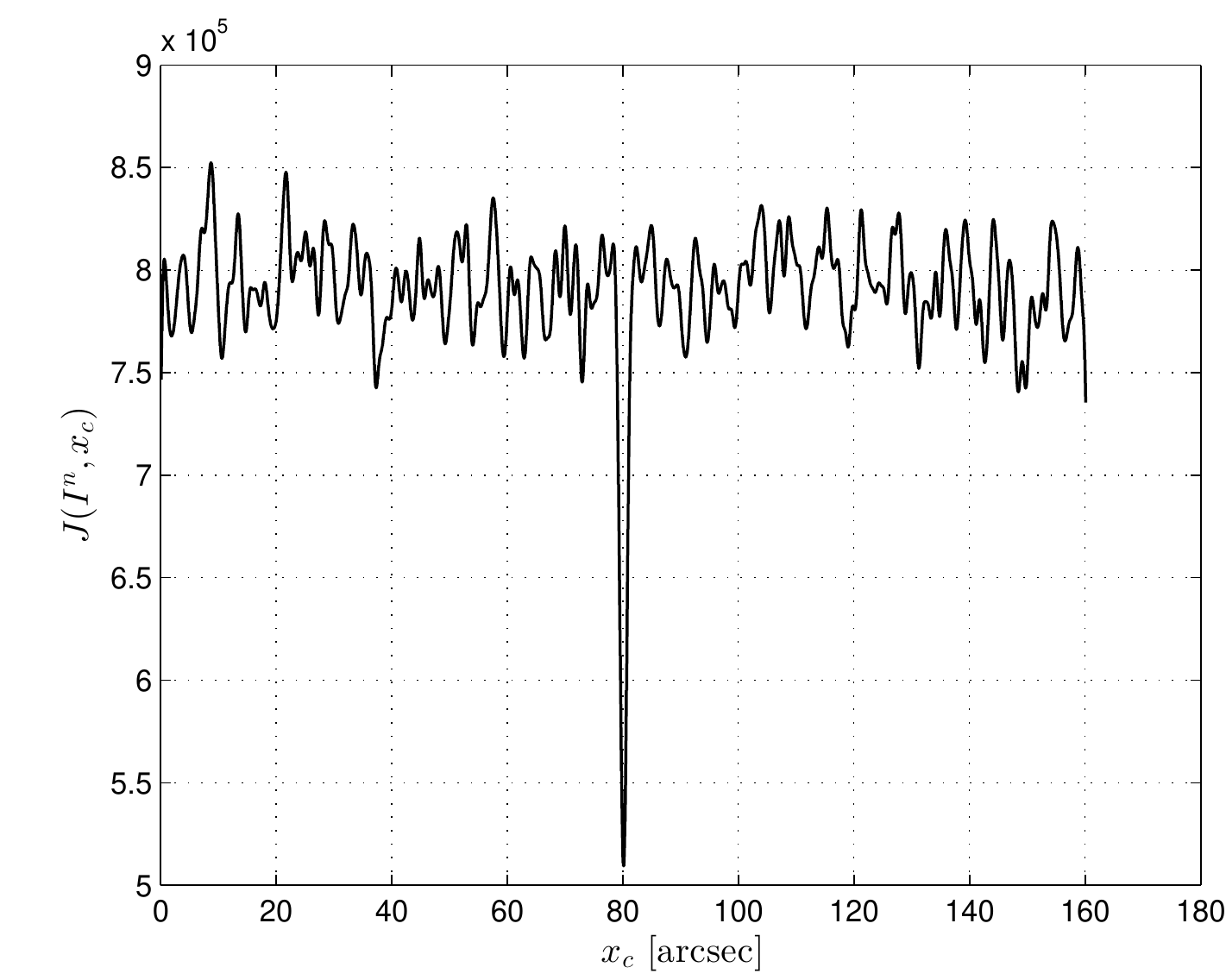}{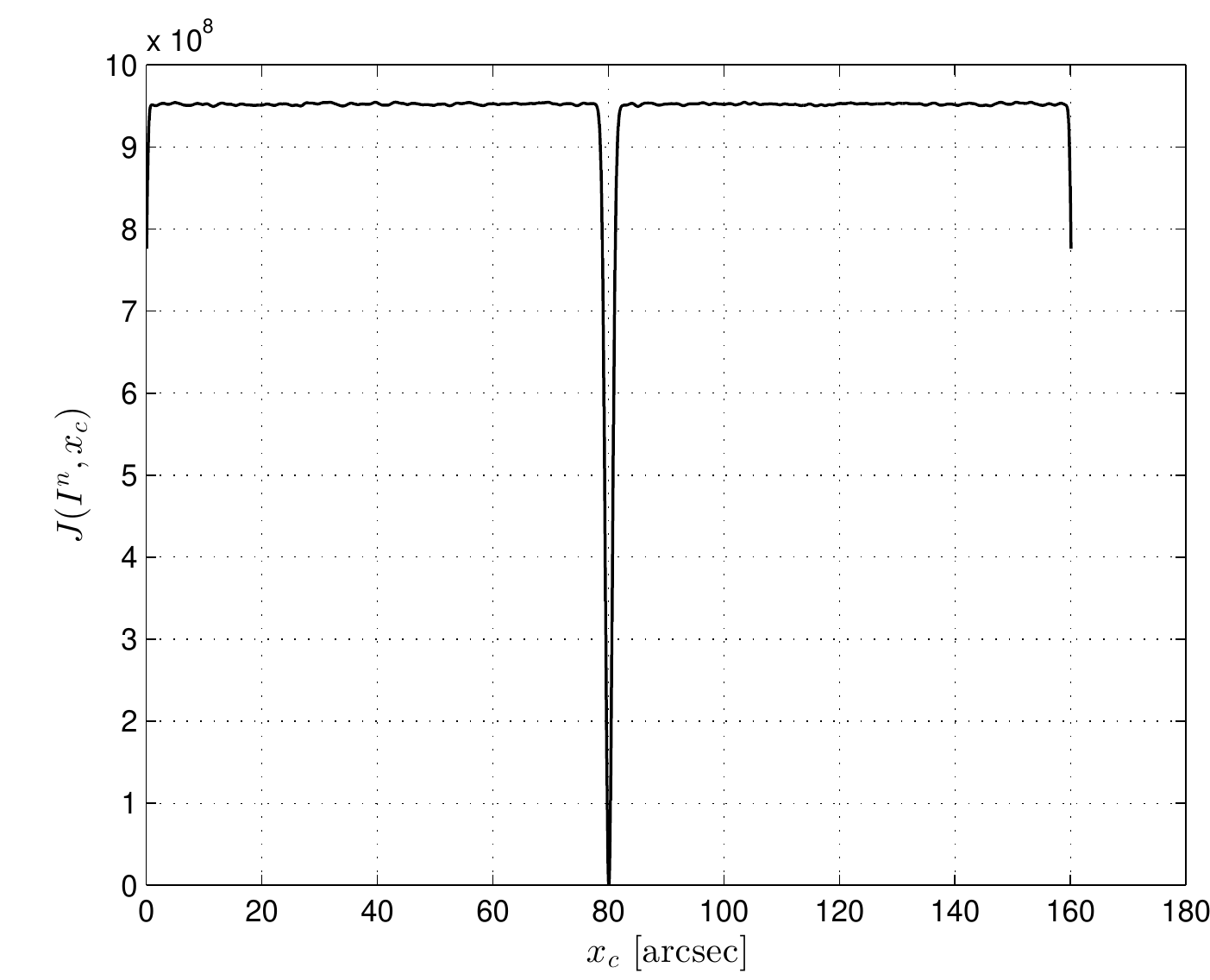}
\caption{Example behavior of the likelihood function $\ln L(I^n,x_c)$
  (upper row) and LS function $J(I^n,x_c)$ (lower row) for two
  particular realizations with low SNR= 12 (left column) and high SNR=
  230 (right column), for a target centered at $x_c=80$~arcsec (for a
  pixel size of $\Delta x=0.2$~arcsec). All the other parameters are
  those described in Section~\ref{subsec_empirical}. The figures show
  (particularly the ones at low SNR) the non-linear nature of both
  objective functions.}\label{fig_behave}
\end{figure}

To refine the relative performance analysis presented in
Figure~\ref{fig4}, Figure~\ref{fig:inter} shows the feasible range
(predicted by Theorem \ref{ls_performances_bounds}) of performance gap
considering the critical $\delta$ obtained in Figure~\ref{fig2}.
We report four cases, from very low to very high SNR
regimes, to illustrate the trends. From this figure, we can see that
the deviations from the nominal value are quite significant for the
low SNR regime, and that, from this perspective, the range obtained
from Theorem \ref{ls_performances_bounds} is not sufficiently small to
conclude about the goodness of the LS estimator in this context. On
the other hand, in the high SNR regime, the nominal comparison can be
considered quite precise.

The results of the previous paragraph motivate an empirical analysis
to estimate the performance of the LS estimator empirically from the
data, with the goal of resolving the low SNR regime illustrated in
Figure~\ref{fig:inter}. For this purpose, $1\,000$ realizations were
considered for all the SNR regimes and pixel sizes, and the
performance of the LS estimator was computed using the empirical
MSE. We used a large number of samples to guarantee convergence to the
true MSE error as a consequence of the law of large numbers
\citep{graydavi10}. Remarkably, we observe in all cases that the
estimated performance matches quite tightly the nominal
$\sigma_{LS}^2$ characterized by Theorem \ref{ls_performances_bounds}.
We illustrate this in Figure~\ref{fig7}, which considers the most
critical low SNR regime\footnote{Similar results were obtained in the
  other regimes, not reported here for the sake of brevity.}.
Consequently, from this numerical analysis, we can resolve the
ambiguity present in the low SNR regime, and conclude that the
comparison with the nominal result reported in Figure~\ref{fig4}, and
the derived conclusion about the LS estimator in the low and high SNR
regimes, can be considered valid. Our simulations also show that the
LS estimator is unbiased. Overall, these results suggests that Theorem
\ref{ls_performances_bounds} could be improved, perhaps by imposing
milder sufficient conditions, in order to prove that $\sigma_{LS}^2$
is indeed a precise indicator of the MSE of the LS estimator at any
SNR regime.

For completeness, we show in Figure~\ref{fig_behave} the behavior of
the log-likelihood function (computed using equation~(\ref{log-like}))
and the LS function $J(I^n,x_c)$ (computed using
equation~(\ref{eq_subsec_mse_of_LS_1})) for our two extreme SNR cases
of our numerical simulations, namely SNR= 12 and SNR= 230. In these
figures the true astrometric position is at $x_c=80$~arcsec (= 400~pix
with $\Delta x$ = 0.2~arcsec). These figures clearly show
(particularly the one at low SNR) the non-linear nature of both
objective functions.

\section{Summary and Final Remarks}
\label{final}

Our work provides results to characterize the performance of the
widely used LS estimator as applied to the problem of astrometry when
derived from digital discrete array detectors.  The main result
(Theorem \ref{ls_performances_bounds}) provides in closed-form a
nominal value ($\sigma^2_{LS}(n)$), and a range around it, for the MSE
of the LS estimator as well as its bias. From the predicted nominal
value, we analyzed how efficient is the LS estimator in comparison
with the MV CR bound.  In particular, we show that the LS estimator is
efficient in the regime of low SNR (a point source with a weak
signal), in the sense that it approximates very closely the CR
bound. On the other hand, we show that at high SNR there is a
significant gap in the performance of the LS estimator with respect to
the MV bound. We believe that this sub-optimal behavior is caused by
the Poissonian nature of the detection process, in which the variance
per pixel increases as the signal itself. Since the LS method is very
sensitive to outliers, the large excursions caused by the large pixel
intensity variance at high SNR make the LS method less efficient (from
the point of view of its MSE), than allowed by the CR bound. These
performance analyses complement and match what has been observed in
photometric estimation, where only in the low SNR regime the LS
estimator has been shown to asymptotically achieve the CR bound.

\begin{figure}[h]
\centering
\includegraphics[width=0.7\textwidth]{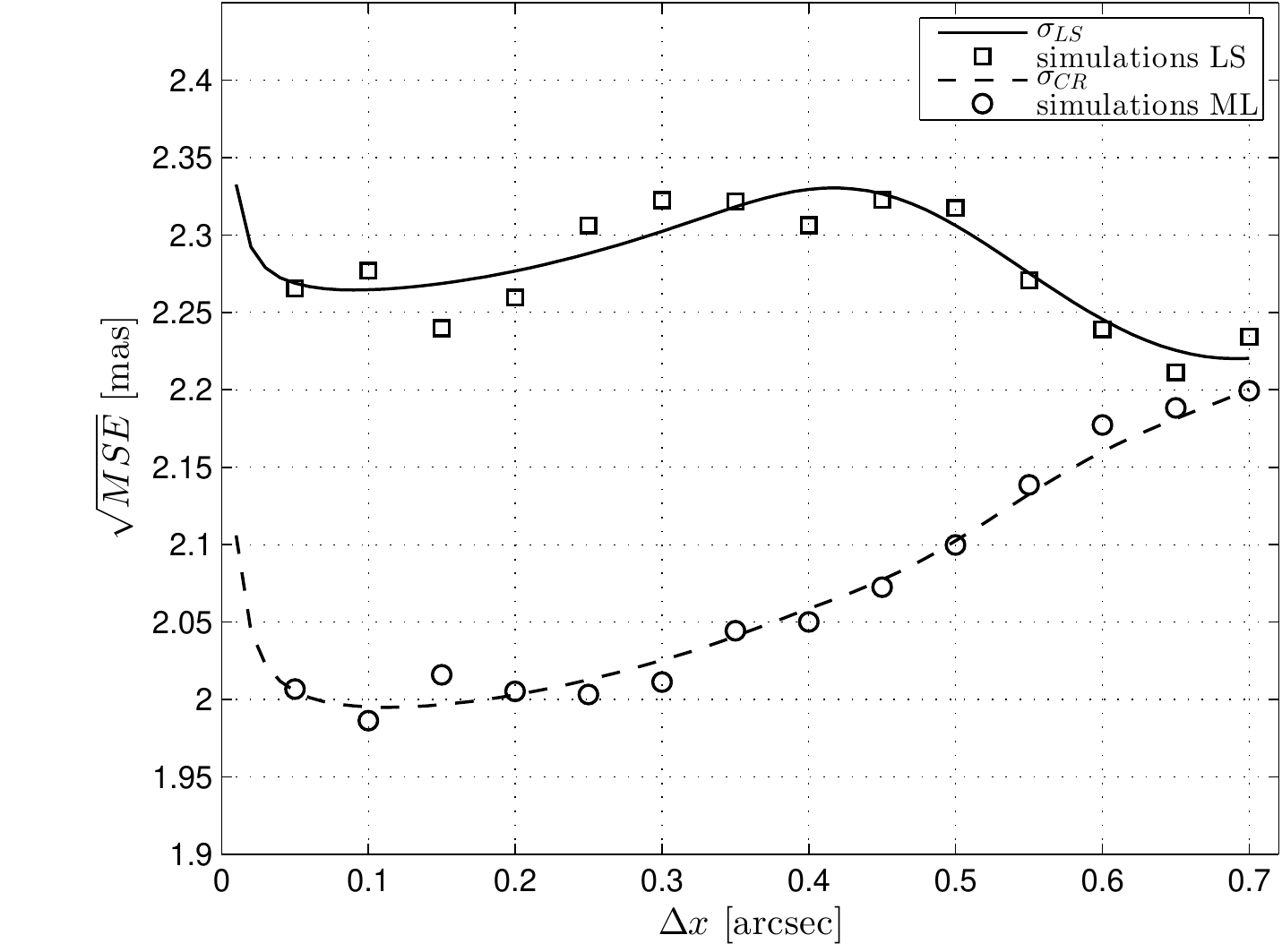}
\caption{Comparison of the $\sqrt{MSE}$ derived from numerical
  simulations using the ML (open circles, equation~(\ref{log-like}))
  and the LS method (open squares,
  equation~(\ref{eq_subsec_mse_of_LS_1})) for a high SNR= 230 (see,
  e.g., right column of Figure~\ref{fig_behave}), where the optimality
  loss (performance gap, $\sim$15\% in this case) of the LS method
  (Proposition~\ref{pro_gap_CR_LS_HSNR}) in this regime is clearly
  seen. The solid line is the nominal $\sigma_{LS}(n)$ derived from
  our theorem (equation~(\ref{nominal})), while the dashed line is the
  CR limit, $\sigma_{CR}$, given by equation~(\ref{fi_astrometry}). As
  we have shown (Section~(\ref{subsec_achie})), the CR limit can not
  be reached in our astrometric setting, but our ML simulations (open
  circles) show that they can follow very closely this limit (see also
  \citet[Table 3]{2013mendez}). A detailed analytical study of the
  optimality of the ML method will be presented in a forthcoming
  paper.} \label{fig_mlvsls}
\end{figure}

While our results are valid for an idealized linear (one-dimensional)
array detector where intra-pixel response changes are neglected, and
where flat-fielding is achieved with very high accuracy, our findings
should motivate the exploration of alternative estimators in the high
SNR observational regime. Regarding this last point, we note that an
inspection of \citet[Table 3]{2013mendez} suggests that either a
(Poisson variance-) weighted LS or a ML approach do not exhibit this
loss of optimality at high-SNR, and should be preferred to the
unweighted LS analyzed in this paper. This effect is clearly
  illustrated in Figure~\ref{fig_mlvsls}, where we present a
  comparison between the standard deviation of the LS and ML methods
  derived from numerical simulations in the very high SNR regime
  (where the gap between CR and the LS is most significant). Motivated
  by these results, a detailed analysis of the ML method will be
  presented in a forthcoming paper.
  
\section{Acknowledgments}

We are indebted to an anonymous referee that read the draft carefully
and in great detail, providing us with several suggestions and
comments that have improved the legibility of the text
significantly. In particular his/her suggestions have lead to the
introduction of Figures~7, 8 and the corresponding discussion in the
body of the paper.

This material is based on work supported by a grant from
CONICYT-Chile, Fondecyt \# 1151213.  In addition, the work of
J. F. Silva and M. Orchard is supported by the Advanced Center for
Electrical and Electronic Engineering, Basal Project
FB0008. J. F. Silva acknowledges support from a CONICYT-Fondecyt grant
\# 1140840 , and R. A. Mendez acknowledges support from Project
IC120009 Millennium Institute of Astrophysics (MAS) of the Iniciativa
Cient\'{\i}fica Milenio del Ministerio de Econom\'{\i}a, Fomento y
Turismo de Chile. R.A.M also acknowledges ESO/Chile for hosting him
during his sabbatical-leave during 2014.


\newpage

\appendix

\section{Proof of Proposition \ref{pro_achie_astrometry}}
\label{proof_pro_achie_astrometry}
We use the well-known fact that the CR bound is achieved by an
unbiased estimator, if and only if, the following decomposition holds
\citep[pp. 12]{kendall1999}
\begin{equation}\label{eq_proof_pro_achie_astrometry_1}
	\frac{d \ln  L(I^n;\theta)}{d \theta} = A(\theta,n) \cdot (f(I^n)-\theta), 
\end{equation}
where $L(I^n;\theta)$ is the likelihood of the observation $I^n \in
\mathbb{N}^n$ given $\theta$, $A(\theta,n)$ is a function of $\theta$
and $n$ alone (i.e., it does dot depend on the data), while $f(I^n)$
is a function of the data exclusively (i.e., it does not depend on the
parameter). Furthermore, if the achievability condition in
equation~(\ref{eq_proof_pro_achie_astrometry_1}) is satisfied, then
$A(\theta,n)=\mathcal{I}_{\theta}(n)$, and $f(I^n)$ is an unbiased
estimator of $\theta$ that achieves the CR bound.

The proof follows by contradiction, assuming that
equation~(\ref{eq_proof_pro_achie_astrometry_1}) holds. First using
equation~(\ref{eq_pre_5}), we have that:
\begin{eqnarray}\label{eq_proof_pro_achie_astrometry_2}
	\frac{d \ln L(I^n; x_c)}{d x_c} = \sum_{i=1}^{n} \left[
          \frac{I_i}{\lambda_i(x_c)} \cdot \frac{d \lambda_i(x_c)}{d x_c} -
          \frac{d \lambda_i(x_c)}{d x_c} \right] = \sum_{i=1}^{n}
        \frac{I_i}{\lambda_i(x_c)} \cdot \frac{d \lambda_i(x_c)}{d x_c},
\end{eqnarray}
the last equality comes from the fact that $\sum_{i=1}^n g_i(x_c)=1$
from the assumption in equation~(\ref{eq_pre_4}).  Then replacing
equations~(\ref{eq_proof_pro_achie_astrometry_2})
and~(\ref{fi_astrometry}) in
equation~(\ref{eq_proof_pro_achie_astrometry_1}):
\begin{eqnarray}\label{eq_proof_pro_achie_astrometry_3}
	f(I^n)=\sum_{i=1}^n \frac{\frac{I_i}{\tilde{F} g_i(x_c)+\tilde{B}_i} \cdot
        \frac{d g_i(x_c)}{d x_c}} { \tilde{F} \sum_{i=1}^n
          \frac{1}{\tilde{F} g_i(x_c)+\tilde{B}_i} \cdot \left( \frac{d
            g_i(x_c)}{d x_c} \right)^2} +x_c,
\end{eqnarray}
which contradicts the assumption that $f(I^n)$ should be a function of
the data alone. Furthermore, if we consider the extreme high SNR
regime, where $\tilde{F} g_i(x_c) \gg \tilde{B}_i$ for all $i$, and
the low SNR regime, where $\tilde{F} g_i(x_c) \ll \tilde{B}_i$ for all
$i$, it follows that:
\begin{equation}\label{eq_proof_pro_achie_astrometry_4}
	f(I^n)=\sum_{i=1}^n I_i \cdot \frac{d \ln g_i(x_c)}{d x_c} \cdot
        \left[ \tilde{F} \sum_{i=1}^n \left( \frac{d g_i(x_c)}{d x_c}
          \right)^2/g_i(x_c) \right]^{-1} +x_c
          \end{equation}
          and \begin{equation}
          f(I^n)=\sum_{i=1}^n \frac{I_i}{\tilde{B}_i} \cdot \frac{d
          g_i(x_c)}{d x_c} \cdot \left[ \tilde{F} \sum_{i=1}^n \left(
          \frac{d g_i(x_c)}{d x_c} \right)^2/ \tilde{B}_i \right]^{-1}
        +x_c,
		\end{equation}
respectively. Therefore a contradiction remains even in these extreme
SNR regimes.

\newpage

\section{Proof of Theorem \ref{ls_performances_bounds}}
\label{proof_ls_performances_bounds}
	The approach of \citet{dsp2013} uses the fact that the
        objective function $J(I^n,\alpha)$ in
        equation~(\ref{eq_subsec_mse_of_LS_1}) is two times
        differentiable, which is satisfied in our context.  As a
        short-hand, if we denote by $\hat{x}_c$ the LS estimator
        solution $\tau_{LS}(I^n)$, then the first order necessary
        condition for a local optimum requires that $J'(I^n,\hat{x}_c)
        \equiv \frac{d J(I^n,\alpha)}{d
          \alpha}|_{\alpha=\hat{x}_c}=0$. The other key assumption in
        \citet{dsp2013} is that $\hat{x}_c$ is in a close neighborhood
        of the true value $x_c$. In our case, this has to do with the
        quality of the pixel-based data used for the inference, which
        we assume it offers a good estimation of the position (see,
        e.g., \citet{king1983accuracy,stone1989}).  Then using a first
        order Taylor expansion of $J'(I^n,\hat{x}_c)$ around $x_c$,
        the following key approximation can be adopted \citep[their
          equation~(4)]{dsp2013}:
	\begin{equation}\label{eq_proof_1}
		0= J'(I^n,\hat{x}_c)= J'(I^n,x_c) + (\hat{x}_c- x_c) \cdot
                J''(I^n,x_c) \Leftrightarrow (x_c - \hat{x}_c) =
                \frac{J'(I^n,x_c)}{J''(I^n,x_c)},
	\end{equation}
	where $J''(I^n,x_c)\equiv \frac{d^2 J(I^n,\alpha)}{d^2
          \alpha}|_{\alpha=x_c}$.  If we consider $I^n \sim f_{x_c}$,
        then from equation~(\ref{eq_proof_1}):
	\begin{equation}\label{eq_proof_2}
	 	x_c - \tau_{LS}(I^n) =
                \frac{J'(I^n,x_c)}{J''(I^n,x_c)}.
	\end{equation}
	The second step in the approximation proposed by
        \citet{dsp2013} is to bound $\frac{J'(I^n,x_c)}{J''(I^n,x_c)}$
        by $\frac{J'(I^n,x_c)}{ \mathbb{E}_{I^n\sim f_{x_c}} \left\{
          J''(I^n,x_c) \right\}}$.  For that we introduce the residual
        variable $W(I^n,\alpha)$ where $J''(I^n,x_c)=
        \mathbb{E}_{I^n\sim f_{x_c}} \left\{ J''(I^n,x_c) \right\}
        (1+W(I^n, x_c))$.  Using the fact that $W(I^n,\alpha)$ is
        bounded almost surely (see Remark~\ref{rem_1}, and
        Section~\ref{anal_bound}):
	\begin{eqnarray}\label{eq_proof_3}
	 \left| \frac{J'(I^n,x_c)}{ \mathbb{E}_{I^n\sim f_{x_c}}
            \left\{ J''(I^n,x_c) \right\}} -
          \frac{J'(I^n,x_c)}{J''(I^n,x_c)} \right|
           &=& \left|
          \frac{J'(I^n,x_c)}{ \mathbb{E}_{I^n\sim f_{x_c}} \left\{
            J''(I^n,x_c) \right\}} \cdot \left[1-\frac{1}{1+W(I^n,
              x_c)} \right] \right| \nonumber\\ &\leq& \left|
          \frac{J'(I^n,x_c)}{ \mathbb{E}_{I^n\sim f_{x_c}} \left\{
            J''(I^n,x_c) \right\}} \right| \cdot \max_{w \in (-\delta,
            \delta)} \left|1-\frac{1}{1+w} \right| \nonumber\\ &\leq&
          \frac{ \left| J'(I^n,x_c) \right|}{ \mathbb{E}_{I^n\sim
              f_{x_c}} \left\{ J''(I^n,x_c) \right\}} \cdot
          \frac{\delta}{1-\delta},
	\end{eqnarray}
	 the last step uses the fact that $\mathbb{E}_{I^n\sim
           f_{x_c}} \left\{ J''(I^n,x_c) \right\}\geq 0$ (see Remark
         \ref{rem_1}).  On the other hand, {\em Jensen's inequality}
         \citep{cover_2006} guarantees that:
	\begin{eqnarray}\label{eq_proof_4}
	 	\left|\frac{ \mathbb{E}_{I^n\sim f_{x_c}}  \left\{J'(I^n,x_c) \right\}}{ \mathbb{E}_{I^n\sim f_{x_c}}  \left\{ J''(I^n,x_c) \right\}} - \mathbb{E}_{I^n\sim f_{x_c}}  \left\{ \frac{J'(I^n,x_c)}{J''(I^n,x_c)} \right\} \right|
	 	  &\leq&  \mathbb{E}_{I^n\sim f_{x_c}}  \left\lbrace\left| \frac{J'(I^n,x_c)}{ \mathbb{E}_{I^n\sim f_{x_c}}  \left\{ J''(I^n,x_c) \right\}} - \frac{J'(I^n,x_c)}{J''(I^n,x_c)} \right|\right\rbrace \nonumber\\
		 &\leq&  \frac{ \mathbb{E}_{I^n\sim f_{x_c}} \left\lbrace\left| J'(I^n,x_c) \right|\right\rbrace}{ \mathbb{E}_{I^n\sim f_{x_c}}  \left\{ J''(I^n,x_c) \right\}} \cdot  \frac{\delta}{1-\delta},
	\end{eqnarray}
	where the last inequality comes from
        equation~(\ref{eq_proof_3}).  Then we use that $J'(I^n,x_c)=
        2 \cdot \sum_{i=1}^n (\lambda_i(x_c)-I_i) \cdot \frac{d\lambda_i(x_c)}{d
          x_c}$, and consequently $\mathbb{E}_{I^n\sim f_{x_c}}
        \left\{ J'(I^n,x_c) \right\}=0$.  Then from
        equations~(\ref{eq_proof_4}) and (\ref{eq_proof_2}), we have
        that:
	\begin{equation}\label{eq_proof_5}
		\left| x_c - \mathbb{E}_{I^n\sim f_{x_c}} \left\{
                \tau_{LS}(I^n) \right\} \right| \leq \frac{
                  \mathbb{E}_{I^n\sim f_{x_c}} \left\lbrace\left| J'(I^n,x_c)
                  \right|\right\rbrace}{ \mathbb{E}_{I^n\sim f_{x_c}} \left\{
                  J''(I^n,x_c) \right\}} \cdot
                \frac{\delta}{1-\delta},
	\end{equation}
	which leads to equation~(\ref{eq_subsec_mse_of_LS_2a}).
	
	Concerning the MSE, from the hypothesis on $W(I^n,x_c)$ we have that: 
%
	\begin{equation}
		\frac{J'(I^n,x_c)^2}{ \mathbb{E}_{I^n\sim f_{x_c}}
                  \left\{ J''(I^n,x_c) \right\} ^2 (1+\delta)^2} 
                  \leq
                \left( \frac{J'(I^n,x_c)}{J''(I^n,x_c)} \right)^2
                 \leq
                \frac{J'(I^n,x_c)^2}{ \mathbb{E}_{I^n\sim f_{x_c}}
                  \left\{ J''(I^n,x_c) \right\} ^2 (1-\delta)^2},\label{eq_proof_6}
	\end{equation}
	almost surely.  Then taking the expected value in
        equation~(\ref{eq_proof_6}) and using
        equation~(\ref{eq_proof_2}) for the central term, it follows
        that:
%
	\begin{equation}
		\frac{ \mathbb{E}_{I^n\sim f_{x_c}} \left\{
                  J'(I^n,x_c)^2 \right\}}{ \mathbb{E}_{I^n\sim
                    f_{x_c}} \left\{ J''(I^n,x_c) \right\}
                  ^2 (1+\delta)^2}
                  \leq \mathbb{E}_{I^n\sim f_{x_c}}
                \left\lbrace\left(x_c - \tau_{LS}(I^n) \right)^2\right\rbrace\nonumber\\
                 \leq
                \frac{\mathbb{E}_{I^n\sim f_{x_c}} \left\{
                  J'(I^n,x_c)^2 \right\}}{ \mathbb{E}_{I^n\sim
                    f_{x_c}} \left\{ J''(I^n,x_c) \right\}
                  ^2 (1-\delta)^2},\label{eq_proof_7}
	\end{equation}
	which concludes the result.
%

\newpage

\section{Proof of Proposition \ref{pro_gap_CR_LS_HSNR}}
\label{proof_pro_gap_CR_LS_HSNR}
Recalling from equation~(\ref{eq_pre_3}) that
$g_i(x_c)=\int_{x_i-\frac{\Delta x}{2}}^{x_i+\frac{\Delta x}{2}}\phi
(x-x_c,\sigma) \, dx$ and assuming a Gaussian PSF of the form
$\phi(x,\sigma)= \frac{1}{\sqrt{2\pi} \, \sigma}
\exp{\left(\frac{-x^2}{2 \sigma^2}\right)}$ (see
Section~\ref{subsec_empirical}) by the mean value theorem and the
hypothesis of small pixel ($\Delta x/\sigma \ll 1$), it is possible to
state that:
\begin{equation}\label{eq_proof_gap_1}
g_i(x_c)\approx\phi(x_i-x_c,\sigma) \cdot \Delta x,
\end{equation}
and then we have that:
\begin{eqnarray}\label{eq_proof_gap_2}
\frac{dg_i(x_c)}{dx_c}&\approx&
\frac{(x_i-x_c)}{\sigma^2} \cdot \frac{1}{\sqrt{2\pi} \, \sigma}\exp
\left(\frac{-(x_i-x_c)^2}{2\sigma^2}\right) \cdot \Delta x \nonumber\\ &=&
\frac{(x_i-x_c)}{\sigma^2} \cdot \phi(x_i-x_c,\sigma) \cdot \Delta x.
\end{eqnarray}
With the above approximation we have that:
\begin{eqnarray} \label{eq_proof_gap_3}
\sum_{i=1}^{n}\left(\frac{dg_i(x_c)}{dx_c}\right)^2g_i(x_c) 
&\approx&
\sum_{i=1}^n\left(\frac{(x_i-x_c)}{\sigma^2} \cdot \phi(x_i-x_c,\sigma) \cdot \Delta
x\right)^2 \cdot \phi(x_i-x_c,\sigma) \cdot \Delta x \nonumber\\
&\approx&
\frac{\Delta x^2}{2\sqrt{3}\pi\sigma^6} \cdot \sum_{i=1}^n(x_i-x_c)^2\frac{1}{\sqrt{2\pi}\frac{\sigma}{\sqrt{3}}}\exp\left(\frac{-(x_i-x_c)^2}{2\left(\frac{\sigma}{\sqrt{3}}\right)^2}\right) \cdot \Delta x. \label{eq_proof_gap_3b}
\end{eqnarray}
The term inside the summation in equation~(\ref{eq_proof_gap_3b}) can
be approximated by an integral due to the small-pixel
hypothesis. Assuming that the source is well sampled by the detector
(see Section \ref{sub_sec_astro_photo}, equation~(\ref{eq_pre_4})) we
can obtain that:
\begin{eqnarray}\label{eq_proof_gap_4}
\sum_{i=1}^{n}\left(\frac{dg_i(x_c)}{dx_c}\right)^2g_i(x_c) 
&\approx&
\frac{\Delta x^2}{2\sqrt{3}\pi\sigma^6}
\int_{-\infty}^{+\infty}(x-x_c)^2\frac{1}{\sqrt{2\pi}\frac{\sigma}{\sqrt{3}}}\exp\left(\frac{-(x-x_c)^2}{2\left(\frac{\sigma}{\sqrt{3}}\right)^2}\right)dx
\\ &=& \frac{\Delta x^2}{2\sqrt{3}\pi\sigma^6}\cdot \frac{\sigma^2}{3}
= \frac{\Delta x^2}{6\sqrt{3}\pi\sigma^4} \label{eq_proof_gap_4b},
\end{eqnarray}
where equation~(\ref{eq_proof_gap_4b}) follows from the fact that the
term inside the integral in equation~(\ref{eq_proof_gap_4})
corresponds to the second moment of a normal random variable of mean
$x_c$ and variance $\frac{\sigma^2}{3}$.

By the same set of arguments used to approximate
$\sum_{i=1}^{n}\left(\frac{dg_i(x_c)}{dx_c}\right)^2 g_i(x_c)$ in
equation~(\ref{eq_proof_gap_4b}), we have that:
\begin{eqnarray}\label{eq_proof_gap_5}
\sum_{i=1}^{n}\left(\frac{dg_i(x_c)}{dx_c}\right)^2 
&\approx&
\sum_{i=1}^n\left(\frac{(x_i-x_c)}{\sigma^2} \cdot \phi(x_i-x_c,\sigma) \cdot \Delta
x\right)^2 \nonumber\\
&\approx& \frac{\Delta
  x}{2\sqrt{\pi}\sigma^5} \cdot \sum_{i=1}^n(x_i-x_c)^2\frac{1}{\sqrt{2\pi}\frac{\sigma}{\sqrt{2}}}\exp\left(\frac{-(x_i-x_c)^2}{2\left(\frac{\sigma}{\sqrt{2}}\right)^2}\right) \cdot \Delta x\nonumber\\
&\approx&
 \frac{\Delta x}{2\sqrt{\pi}\sigma^5} \int_{-\infty}^{+\infty}(x-x_c)^2\frac{1}{\sqrt{2\pi}\frac{\sigma}{\sqrt{2}}}\exp\left(\frac{-(x-x_c)^2}{2\left(\frac{\sigma}{\sqrt{2}}\right)^2}\right)dx   \label{eq_proof_gap_5b}\\
&\approx& \frac{\Delta x}{2\sqrt{\pi}\sigma^5}\cdot \frac{\sigma^2}{2}
=\frac{\Delta x}{4\sqrt{\pi}\sigma^3}\label{s_aprox},
\end{eqnarray}
where equation~(\ref{s_aprox}) follows from the fact that the term
inside the integral in equation~(\ref{eq_proof_gap_5b}) is the second
moment of a normal random variable of mean $x_c$ and variance
$\frac{\sigma^2}{\sqrt{2}}$.

Finally, for $\sum_{i=1}^n\left(\frac{dg_i(x_c)}{dx_c}\right)^2 \cdot
\frac{1}{g_i(x_c)}$ we proceed again in the same way, namely:
\begin{eqnarray} \label{eq_proof_gap_6}
\sum_{i=1}^{n}\left(\frac{dg_i(x_c)}{dx_c}\right)^2\frac{1}{g_i(x_c)}&\approx&
\sum_{i=1}^n\left(\frac{(x_i-x_c)}{\sigma^2} \cdot \phi(x_i-x_c,\sigma) \cdot \Delta
x\right)^2\frac{1}{\phi(x_i-x_c,\sigma) \cdot \Delta x}\nonumber\\
&\approx&
\frac{1}{\sigma^4}\sum_{i=1}^n(x_i-x_c)^2\frac{1}{\sqrt{2\pi}\sigma}\exp\left(\frac{-(x_i-x_c)^2}{2\sigma^2}\right)\Delta x\nonumber\\
&\approx&
\frac{1}{\sigma^4} \int_{-\infty}^{+\infty}\frac{(x-x_c)^2}{\sqrt{2\pi}\sigma}\exp\left(\frac{-(x-x_c)^2}{2\sigma^2}\right)dx \label{f_aprox}\\
&\approx& \frac{1}{\sigma^4}\cdot \sigma^2
= \frac{1}{\sigma^2},\label{t_aprox}
\end{eqnarray}
where equation~(\ref{t_aprox}) follows from the fact that the term
inside the integral in equation~(\ref{f_aprox}) corresponds to the
second moment of a normal random variable of mean $x_c$ and variance
$\sigma^2$.

Then adopting equations~(\ref{eq_proof_gap_4b}), (\ref{s_aprox}) and
(\ref{t_aprox}) in equations~(\ref{eq_subsec_mse_of_LS_4}) and
(\ref{fi_astrometry}), respectively,
and assuming a uniform background underneath the PSF, we have that:
\begin{eqnarray}\label{eq_proof_gap_7}
\sigma_{LS}^2
&=&\frac{\sum_{i=1}^n
  \left(\frac{dg_i(x_c)}{dx_c}\right)^2g_i(x_c)}{\tilde{F}\left(\sum_{i=1}^n
  \left(\frac{dg_i(x_c)}{dx_c}\right)^2\right)^2}+\frac{\tilde{B}}{\tilde{F}^2\sum_{i=1}^n
  \left(\frac{dg_i(x_c)}{dx_c}\right)^2}\nonumber\\
&\approx&\frac{\frac{\Delta x^2}{6\sqrt{3}\pi\sigma^4}}{\tilde{F}\left(\frac{\Delta x}{4\sqrt{\pi}\sigma^3}\right)^2}+\frac{\tilde{B}}{\tilde{F}^2\frac{\Delta x}{4\sqrt{\pi}\sigma^3}}\nonumber\\
&=& \frac{\sigma^2}{\tilde{F}}\left[\frac{8}{3\sqrt{3}}+4\sqrt{\pi}\frac{1}{\frac{\Delta x}{\sigma}}\frac{\tilde{B}}{\tilde{F}}\right]
\approx \frac{\sigma^2}{\tilde{F}}\frac{8}{3\sqrt{3}} \label{ap_ls_f} \label{eq_proof_gap_7b}
\end{eqnarray}
and,
\begin{equation}\label{eq_proof_gap_8}
\sigma_{CR}^2=\frac{1}{\sum_{i=1}^n\frac{\left(\tilde{F}\frac{dg_i(x_c)}{dx_c}\right)^2}{\tilde{F}g_i(x_c)+\tilde{B}}}
=\frac{1}{\tilde{F}\sum_{i=1}^n\left(\frac{dg_i(x_c)}{dx_c}\right)^2\frac{1}{g_i(x_c)}}
\approx\frac{\sigma^2}{\tilde{F}}.
\end{equation}
In the last step in equation~(\ref{eq_proof_gap_7b}) and the last step
in equation~(\ref{eq_proof_gap_8}), we have used the assumption that
$\tilde{F} \gg \tilde{B}$. We note that the last step in
equation~(\ref{eq_proof_gap_8}) corresponds to the second line of
equation~(45) in \citet{2013mendez}.

\newpage




\newpage

\begin{thebibliography}{}

\bibitem[\protect\citeauthoryear{Adorf}{Adorf}{1996}]{adorf1996}
  Adorf, H.-M. 1996, in Astronomical Data Analysis Software and
  Systems V, Vol.  101, 13

\bibitem[\protect\citeauthoryear{Alard \& Lupton}{Alard \&
    Lupton}{1998}]{alard1998method} Alard, C., \& Lupton, R.~H. 1998,
  \aj, 503, 325

\bibitem[\protect\citeauthoryear{Auer \& van~Altena}{Auer \&
    van~Altena}{1978}]{euer1978} Auer, L., \& van~Altena, W. 1978,
  \aj, 83, 531

\bibitem[Bastian(2004)]{bastian2004} Bastian, U. 2004. GAIA technical
  note, GAIA-C3-TN-ARI-BAS-020
  (http://www.cosmos.esa.int/web/gaia/public-dpac-documents). Available
  at
  http://www.rssd.esa.int/SYS/docs/ll\_transfers/project=PUBDB\&id=2939027.pdf
  (last accessed on April 2015).

\bibitem[Becker et al.(2007)]{becket07} Becker, A.~C., Silvestri,
  N.~M., Owen, R.~E., Ivezi{\'c}, {\v Z}., \& Lupton, R.~H.\ 2007,
  \pasp, 119, 1462

\bibitem[\protect\citeauthoryear{Bendinelli et~al.}{Bendinelli
    et~al.}{1987}]{bendinelli1987} Bendinelli, O., Parmeggiani, G.,
  Piccioni, A., \& Zavatti, F. 1987, \aj, 94, 1095

\bibitem[\protect\citeauthoryear{Cameron et~al.}{Cameron
  et~al.}{2006}]{cameron2006fast}
Cameron, A.~C., et~al. 2006, \mnras,
  373, 799

\bibitem[\protect\citeauthoryear{Chromey}{Chromey}{2010}]{chromey2010measure}
Chromey, F.~R. 2010, To measure the sky: an introduction to observational
  astronomy (Cambridge University Press)

\bibitem[\protect\citeauthoryear{Cover \& Thomas}{Cover \&
  Thomas}{2006}]{cover_2006}
Cover, T.~M.,  \& Thomas, J.~A. 2006, Elements of Information Theory (second
  ed.) (Wiley Interscience, New York)

\bibitem[\protect\citeauthoryear{Cram{\'e}r}{Cram{\'e}r}{1946}]{cramer1946cont%
ribution}
Cram{\'e}r, H. 1946, Scandinavian Actuarial Journal, 1946, 85

\bibitem[\protect\citeauthoryear{Fessler}{Fessler}{1996}]{fessler1996}
Fessler, J.~A. 1996, Image Processing, IEEE Transactions on, 5, 493

\bibitem[Gawiser et al.(2006)]{gawiet06} Gawiser, E., van Dokkum,
  P.~G., Herrera, D., et al.\ 2006, \apjs, 162, 1

\bibitem[\protect\citeauthoryear{{Gilliland}}{{Gilliland}}{1992}]{gilligand199%
  2} {Gilliland}, R.~L. 1992, in Astronomical Society of the Pacific
  Conference Series, Vol.~23, Astronomical CCD Observing and Reduction
  Techniques, ed.  S.~B. {Howell}, 68

\bibitem[Freyhammer et al.(2001)]{freyetal01} Freyhammer, L.~M.,
  Andersen, M.~I., Arentoft, T., Sterken, C., \& N{\o}rregaard,
  P.\ 2001, Experimental Astronomy, 12, 147

\bibitem[\protect\citeauthoryear{Gray \& Davisson}{Gray \&
    Davisson}{2010}]{graydavi10} Gray, R.~M., \& Davisson, L.~D. 2010,
  An Introduction to Statistical Signal Processing, Cambridge
  University Press

\bibitem[\protect\citeauthoryear{H{\o}g}{H{\o}g}{2009}]{hog2009}
H{\o}g, E. 2009, Experimental Astronomy, 25, 225

\bibitem[\protect\citeauthoryear{H{\o}g}{H{\o}g}{2011}]{hog2011}
H{\o}g, E. 2011, Baltic Astronomy, 20, 221

\bibitem[\protect\citeauthoryear{Honeycutt}{Honeycutt}{1992}]{honeycutt1992ccd}
Honeycutt, R.~K. 1992, \pasp,
  435

\bibitem[\protect\citeauthoryear{Howell}{Howell}{2006}]{howell2006handbook}
Howell, S.~B. 2006, Handbook of CCD astronomy, Vol.~5 (Cambridge University
  Press)

\bibitem[\protect\citeauthoryear{Jakobsen, Greenfield, \&
  Jedrzejewski}{Jakobsen et~al.}{1992}]{jakobsen1992}
Jakobsen, P., Greenfield, P.,  \& Jedrzejewski, R. 1992, \aap, 253, 329

\bibitem[Kay (1993)]{kay93} Kay, S. M. 1993, Fundamentals of
  Statistical Signal Processing, Volume I: Estimation Theory. Prentice
  Hall, New Jersey.

\bibitem[\protect\citeauthoryear{Kendall et~al.}{Kendall
  et~al.}{1999}]{kendall1999}
Kendall, M., Stuart, A., Ord, J.,  \& Arnold, S. 1999, Vol. 2A: Classical
  inference and the linear model (London [etc.]: Arnold [etc.])

\bibitem[\protect\citeauthoryear{King}{King}{1971}]{king1971}
King, I.~R. 1971, \pasp, 199

\bibitem[\protect\citeauthoryear{King}{King}{1983}]{king1983accuracy}
King, I.~R. 1983, \pasp, 163

\bibitem[\protect\citeauthoryear{Lattanzi}{Lattanzi}{2012}]{lattanzi2012}
Lattanzi, M. 2012, \memsai, 83, 1033

\bibitem[\protect\citeauthoryear{Lee \& van~Altena}{Lee \&
  van~Altena}{1983}]{lee1983theoretical}
Lee, J.-F.,  \& van~Altena, W. 1983, \aj, 88, 1683

\bibitem[\protect\citeauthoryear{Lindegren}{Lindegren}{1978}]{lindegren1978}
Lindegren, L. 1978, in IAU Colloq. 48: Modern Astrometry, Vol.~1, 197

\bibitem[Lindegren(2000)]{lind2000} Lindegren, L. 2000, Gaia technical
  report Gaia-LL-032

\bibitem[\protect\citeauthoryear{{Lindegren}}{{Lindegren}}{2010}]{lindegren201%
  0} {Lindegren}, L. 2010, ISSI Scientific Reports Series, 9, 279

\bibitem[Lupton et al.(2001)]{luptonetal2001} Lupton, R., Gunn, J.~E.,
  Ivezi{\'c}, Z., Knapp, G.~R., \& Kent, S.\ 2001, Astronomical Data
  Analysis Software and Systems X, 238, 269

\bibitem[Lupton(2007)]{lupton2007} Lupton, R.\ 2007, Statistical
  Challenges in Modern Astronomy IV, 371, 160

\bibitem[Mason (2007)]{mason07} Mason, E. 2008, in The 2007 ESO
  Instrument Calibration Workshop, Kaufer, A., \& Kerber, F. (eds.),
  Springer-Verlag, Berlin, 107

\bibitem[\protect\citeauthoryear{Mendez et~al.}{Mendez
  et~al.}{2010}]{2010mendez}
Mendez, R.~A., Costa, E., Pedreros, M.~H., Moyano, M., Altmann, M.,  \&
  Gallart, C. 2010, \pasp,
  122, 853

\bibitem[\protect\citeauthoryear{Mendez, Silva, \& Lobos}{Mendez
  et~al.}{2013}]{2013mendez}
Mendez, R.~A., Silva, J.~F.,  \& Lobos, R. 2013, \pasp, 125, pp. 580

\bibitem[\protect\citeauthoryear{Mendez et~al.}{Mendez
  et~al.}{2014}]{2014mendez}
Mendez, R.~A., Silva, J.~F., Orostica, R.,  \& Lobos, R. 2014, \pasp, 126, 798

\bibitem[\protect\citeauthoryear{Mighell}{Mighell}{2005}]{mighell2005stellar}
Mighell, K.~J. 2005, \mnras, 361,
  861

\bibitem[Najim (2008)]{najim2008} Najim, M. 2008, Modeling, Estimation
  and Optimal Filtering in Signal Processing, Chapter A,
  335-340. Available at
  http://onlinelibrary.wiley.com/book/10.1002/9780470611104, last
  accessed on June 26th, 2015.

\bibitem[\protect\citeauthoryear{Perryman et~al.}{Perryman
  et~al.}{1989}]{1989perryman}
Perryman, M., Jakobsen, P., Colina, L., Lelievre, G., Macchetto, F., Nieto, J.,
   \& Serego~Alighieri, S. 1989, \aap, 215, 195

\bibitem[Pier et al.(2003)]{pieret03} Pier, J.~R., Munn, J.~A.,
  Hindsley, R.~B., et al.\ 2003, \aj, 125, 1559

\bibitem[\protect\citeauthoryear{Rao}{Rao}{1945}]{ra%
dhakrishna1945information}
Rao, C.~R. 1945, Bulletin of the Calcutta Mathematical Society, 37,
  81

\bibitem[\protect\citeauthoryear{Raykar, Kozintsev, \&
    Lienhart}{Raykar et~al.}{2005}]{raykar2005} Raykar, V.~C.,
  Kozintsev, I.~V., \& Lienhart, R. 2005, Speech and Audio Processing,
  IEEE Transactions on, 13, 70

\bibitem[\protect\citeauthoryear{Reffert}{Reffert}{2009}]{reffert2009}
Reffert, S. 2009, \nar, 53, 329

\bibitem[\protect\citeauthoryear{So et~al.}{So et~al.}{2013}]{dsp2013}
So, H., Chan, Y., Ho, K.,  \& Chen, Y. 2013, Signal Processing Magazine, IEEE,
  30, 162

\bibitem[\protect\citeauthoryear{Stetson}{Stetson}{1987}]{stetson1987daophot}
Stetson, P.~B. 1987, \pasp,
  191

\bibitem[\protect\citeauthoryear{{Stone}}{{Stone}}{1989}]{stone1989}
{Stone}, R.~C. 1989, \aj, 97, 1227

\bibitem[\protect\citeauthoryear{{Tyson}}{{Tyson}}{1986}]{tyson1986}
{Tyson}, J.~A. 1986, Journal of the Optical Society of America A, 3, 2131

\bibitem[\protect\citeauthoryear{van~Altena \& Auer}{van~Altena \&
  Auer}{1975}]{vanAltena_1975}
van~Altena, W.,  \& Auer, L. 1975, in Image Processing Techniques in Astronomy
  (Springer), 411

\bibitem[\protect\citeauthoryear{van Altena}{van Altena}{2013}]{vanaltena2013}
van Altena, W.~F. 2013, Astrometry for Astrophysics: Methods, Models, and
  Applications (Cambridge University Press)

\bibitem[\protect\citeauthoryear{Winick}{Winick}{1986}]{Winick:86}
Winick, K.~A. 1986, J. Opt. Soc. Am. A, 3, 1809

\bibitem[\protect\citeauthoryear{Zaccheo et~al.}{Zaccheo
  et~al.}{1995}]{zaccheo1995}
Zaccheo, T., Gonsalves, R., Ebstein, S.,  \& Nisenson, P. 1995, \apj, 439, L43

\end{thebibliography}
\end{document}